\documentclass[12pt]{article}%
\usepackage{amsmath}
\usepackage{amssymb}
\usepackage{graphicx}
\usepackage{url}
\usepackage{subfigure}
\usepackage{color}
\usepackage{geometry}
\usepackage{setspace}
\usepackage{amsfonts}%
\providecommand{\U}[1]{\protect\rule{.1in}{.1in}}
\newtheorem{theorem}{Theorem}

\newtheorem{claim}[theorem]{Claim}

\newtheorem{definition}[theorem]{Definition}

\newtheorem{remark}[theorem]{Remark}

\newenvironment{proof}[1][Proof]{\textbf{#1.} }{\ \rule{0.5em}{0.5em}}

\begin{document}

\title{Game Representations and Extensions of the Shapley Value\thanks{This is a
revised version of \cite{Dubey}. It is a pleasure to thank Ori Haimanko for
several helpful discussions; as well as Bhavook Bhardwaj, Siddharth Chatterjee
and Takahiro Moriya for several useful comments. } }
\author{Pradeep Dubey\thanks{Center for Game Theory, Department of Economics, Stony
Brook University; and Cowles Foundation for Research in Economics, Yale
University}\ \ }
\date{9 January 2024}
\maketitle

\begin{abstract}
We show that any cooperative game can be represented by an assignment of
costly facilities to players, in which it is intuitively obvious how to
allocate the total cost in an equitable manner. This equitable solution turns
out to be the Shapley value of the game, and thus provides as an alternative
justification of the value.

Game representations also open the door for extending the Shapley value to
situations where not all coalitions can form, provided those that can
constitute a \textquotedblleft semi-algebra\textquotedblright; or, more
generally, a \textquotedblleft hierarchy\textquotedblright;\ or, still more
generally, have \textquotedblleft full span\textquotedblright.


\textbf{JEL\ Classification:} C71, C72, D61, D63, D70, D79

\end{abstract}

\section{Introduction}

Consider a (cooperative) game\footnote{Such games are also known as
\textquotedblleft games in characteristic function form\textquotedblright%
\ ($v$ being the characteristic function), or \textquotedblleft games with
side payments\textquotedblright, or \textquotedblleft games with transferable
utility\textquotedblright.} $v:\mathcal{N\longrightarrow}\mathbb{R}$, where
$\mathbb{R}$ denotes the reals and $\mathcal{N}$ the collection of all
coalitions (i.e., non-empty subsets) of the player set $N$. Here $v\left(
S\right)  $ is the payoff that coalition $S$ can guarantee\textit{ }to itself
by full cooperation among its members, no matter what the other players in
$N\diagdown S$ do. Even though --- for reasons of efficiency\footnote{Though
we do not formally need to make this assumption, we have in mind games that
are superadditive, i.e., $v\left(  S\cup T\right)  \geq v\left(  S\right)
+v\left(  T\right)  $ whenever $S\cap T=\emptyset;$ and in this case the
formation of the coalition $N$ makes for efficiency. (See Section
\ref{superadditive subsection} for a detailed discussion.)} --- only the grand
coalition $N$ takes center-stage, and it is $v\left(  N\right)  $ that
ultimately needs to be allocated amongst the players in $N$, the other
coalitions are present in the background and the entire game $v$ is relevant
for this purpose. A leading candidate for the allocation of $v\left(
N\right)  $ is the Shapley value (\cite{Shapley 1951}, \cite{Shapley 1953}) of
the game $v$ and it will be the center of attention of this paper.

We show that any\emph{ }such game $v$ can be \textquotedblleft
represented\textquotedblright\ by an assignment $\left(  \psi,\gamma\right)  $ where

(i) $\psi\left(  n\right)  \subseteq K$ is an assignment of facilities to
$n\in N$;

(ii) $\gamma\left(  k\right)  $ is the constant cost of facility $k\in K,$
that accrues collectively to \emph{any} set of users of $k$.

The operative word here is \textquotedblleft assignment\textquotedblright:
$\psi(n)$ is simply the set of facilities that player $n$ \emph{must }use,
with no choice in the matter. Also, the word \textquotedblleft
cost\textquotedblright\ is used with the explicit understanding that when
$\gamma\left(  k\right)  <0$ (or, $\gamma\left(  k\right)  >0$) it corresponds
to a cost (or, benefit) in the colloquial sense. (However, we invite the
reader to focus on the case $\gamma\left(  k\right)  <0$ for all $k$, which
corresponds to the well-known cost allocation problem in game theory. See
Section \ref{superadditive subsection} )

The assignment $\left(  \psi,\gamma\right)  $ is said to be a
\emph{representation} of $v$ if $v\left(  S\right)  $ is the total cost of the
facilities assigned to players in $S$, for every coaltion $S\in\mathcal{N}.$

The abstract problem, of the allocation of $v\left(  N\right)  $ in the game
$v,$ may thus be recast in concrete terms within the representation $\left(
\psi,\gamma\right)  $ as follows: how \textquotedblleft
should\textquotedblright\ the total cost $\sum_{k\in K}\gamma\left(  k\right)
$ of the facilities be allocated amongst the users in $N$?

To this end, note that each facility $k$ is \emph{user-blind}\ since
$\gamma\left(  k\right)  ,$ being a constant, is independent of the set of
users of $k$ or of any other facility. So if we adopt the $k$%
-\emph{standpoint},\emph{ }i.e., consider the allocation of $\gamma\left(
k\right)  $ from the standpont of facility $k$ alone, a solution comes almost
unbidden to mind: those who do not use facility $k$ should have to bear none
of its cost; furthermore, since $k$ is user-blind, $\gamma\left(  k\right)  $
should be split equally amongst its users\footnote{i.e., among all those in
the set $\psi^{-1}\left(  k\right)  =\left\{  n\in N:k\in\psi\left(  n\right)
\right\}  .$}. Aggregating across $k\in K$ then yields the overall allocation
of $v\left(  N\right)  $ which we shall call the \emph{equitable solution.}

It turns out that the equitable solution is invariant of the representation of
$v$ and coincides with the Shapley value of $v,$ for \emph{all} games $v$ on
$\mathcal{N}$.

Worthy of note is the fact that the equitable solution is based on \emph{just}
the representation, without any explicit consideration of coalitions in
$\mathcal{N}$, leave aside a characteristic function $v$ on $\mathcal{N}$
(though, of course, $v$ is fully used to construct the representation in the
first place.) Thus game representations not only provide an alternative
outlook on the Shapley value, but they also serve to describe it to those who
are not of a mathematical turn of mind.\footnote{The equitable solution of
assignment $\left(  \psi,\gamma\right)  $ was invariably given as being
evident, by a random assortment of lay people that the author posed the
question to: high school students, fashion designers, actors, businessmen,
politicians, \textit{et al}., all of them perfectly innocent of game theory.
Yet their solution was the \textquotedblleft Shapley value\textquotedblright%
\ of the \textquotedblleft cooperative game\textquotedblright\ represented by
$(\psi,\gamma).$}

One may, if of a scholastic turn of mind, explicate two principles behind the
equitable solution, both so intuitively compelling as to require little
justification. The first is \emph{decentralization. }As was said, the
facilities in our model are \textquotedblleft decentralized\textquotedblright%
\ to begin with: at any facility $k,$ the cost $\gamma\left(  k\right)  $ to
its users does not depend on what is happening at the other facilities, i.e.,
who goes to them and what their costs are. It stands to reason that the
allocaton of $\gamma\left(  k\right)  $ should also inherit this property:
those who do not use facility $k$ should be ignored at $k$ and bear none of
the cost $\gamma\left(  k\right)  .$

The second principle is \emph{non-discrimination}: any facility $k$ should
split its cost $\gamma\left(  k\right)  $ equally among all its users since,
being user-blind even at its own site, it has no basis on which to
discriminate between them.

The game representation approach, outlined above, carries over to situations
where several coalitions are deemed infeasible and
\[
v:\mathcal{C\longrightarrow}\mathbb{R}%
\]
is defined on a subdomain $\mathcal{C\subsetneq N}$. We feel that this
generalized model\footnote{The game on curtailed coalitions $\mathcal{C}$ is a
generalization in a different direction from that of Myerson (\cite{Myerson}).
In \cite{Myerson}, one starts with a full-blown game
$v:\mathcal{N\longrightarrow}\mathbb{R}$, along with a \textquotedblleft
communication graph\textquotedblright\ $g$ on players. The two together then
determine another full-blown game $v^{g}:\mathcal{N\longrightarrow}%
\mathbb{R}.$ In contrast, in our model, there is no extension envisaged of the
game $v:\mathcal{C\longrightarrow}\mathbb{R}$ to a full-blown game on
$\mathcal{N}.$ Coalitions in $\mathcal{N\diagdown C}$ are viewed throughout as
infeasible and never enter the picture.
\par
When $\mathcal{C\subsetneq N}$, several authors have referred to
$v:\mathcal{C\longrightarrow}\mathbb{R}$ as a \textquotedblleft partially
defined cooperative game\textquotedblright\ (see, e.g., \cite{Albizuri},
\cite{Housman}, \cite{Masuya}, \cite{Wilson}). One could alternatively think
of $v$ as a \textquotedblleft generalized cooperative game\textquotedblright%
\ since the standard game corresponds to the special case $\mathcal{C=N}$.}
may be of relevance in practical applications. (See Section \ref{Examples} for
some indicative examples, and \cite{Albizuri}, \cite{Housman}, \cite{Masuya},
\cite{Wilson} for further discussion of such games, as well as complementary
approaches to the Shapley value).

If $\mathcal{C}$ is a \textquotedblleft semi-algebra\textquotedblright, or ---
more generally --- if $\mathcal{C}$ is \textquotedblleft
hierarchical\textquotedblright; or --- still more generally --- if
$\mathcal{C}$ satisfies the \textquotedblleft full span\textquotedblright%
\ condition, then we obtain our main result, which is roughly as follows. (See
Theorem \ref{general TU} in Section \ref{Curtailed}, where the terms within
quotes are made precise, and the result is stated formally.)

\bigskip

\textbf{MAIN RESULT: Suppose }$\mathcal{C}$ \textbf{has full span. Then there
exist representations of every }$v:\mathcal{C\longrightarrow}\mathbb{R}$,
\textbf{and the equitable solution is invariant across all representations of
}$v$\textbf{. Furthermore, in the canonical case }$\mathcal{C=N}$,
\textbf{the} \textbf{equitable solution coincides with the Shapley value of
}$v$ \bigskip

Thus the equitable solution constitutes an extension of the Shapley value to
domains $\mathcal{C\subsetneq N}$ that satisfy the full span condition. (We
also show in Theorem \ref{necessity of full span} that the full span condition
on $\mathcal{C}$ is not only sufficient for the conclusion of Theorem
\ref{general TU} but also necessary; and in this sense the result cannot be sharpened.)

In our analysis, the long-forgotten \textquotedblleft Axiom
I\textquotedblright\ from Shapley's 1951 working paper \cite{Shapley 1951}
plays a pivotal role\footnote{The words within parantheses are ours.}:

\bigskip

\textbf{Axiom I: The (Shapley) value (of any game)} \textbf{depends only on
the characteristic function }$v$\textbf{ (that the game} \textbf{induces).}

\bigskip

Axiom I will look nothing short of bizarre to most readers today, because
games have come to be \emph{identified }with the characteristic functions they
induce. Indeed this is so even in a subsequent version \cite{Shapley 1953} of
\cite{Shapley 1951}, published by Shapley himself just two years later; and,
in deference to that tradition, in this paper as well.

However Axiom I does have significant content, which is presumably what led
Shapley to state it in the first place. In practice, games arise within a
specific \emph{context}, such as the allocation of costs among users of shared
facilities (see the articles in \cite{Roth}, especially \cite{Tauman}), of
which our representations are special instances; or of the gains to trade
among participants in a market (\cite{Shapley-Shubik}); or, especially at the
time that Shapley was writing, gains from collusion within a strategic
conflict (\cite{von Neumann}); and so on\footnote{We have just cited a few
\textquotedblleft ancestral\textquotedblright\ papers here, and are far from
doing justice to the full literature.}. If we take the word \textquotedblleft
game\textquotedblright\ to connote the entire context, as is done in common
parlance\footnote{People talk of chess or bridge\ as a game, not of the
underlying extensive form; or of the allocation of the cost of a network among
its heterogeneous users, not of the abstract characteristic function that
emerges from it.}, then Axiom I embodies the substantive assumption that the
value does not depend on the finer economic or engineering structure of the
\textquotedblleft game\textquotedblright, but only on its emergent
characteristic function. In short, for purposes of cost allocation, what
matters are the costs incurred by coalitions, never mind \emph{how} they
incurred it.

This opens up the possibility of \emph{reversing the gaze}\textit{: }rather
than go from any context $\Gamma$ to the characteristic function $v$ that
$\Gamma$ induces, and then figure out the value to players in $v,$ one could
instead start with $v$ and conjure up a special context $\Gamma^{\ast}$ which
induces $v$ and in which the value is intuitively obvious. And this is just
what we do when we conjure up our game representations. Being
\textquotedblleft coalition-free\textquotedblright\ in their structure, they
are --- from a conceptual (though not computational) viewpoint --- an
order-of-magnitude simpler than the coalitional games that they represent,
which is what enables the equitable solution to come to light immediately,
without further ado.

Our main result thus owes everything to the obsolescent Axiom I of Shapley.

The paper is organized as follows. Sections \ref{games}, \ref{assignments},
\ref{representations}, \ref{Shapley Value}, \ref{Curtailed},
\ref{Ext Shapley Value} define the model and state the main result. Section
\ref{Examples} contains some examples of games with a semi-algebra/hierarchy
of feasible coalitions. In Section \ref{Discriminatory Facilities} we give a
heuristic argument as to why user-blind facilities are more reasonable for our
analysis compared to \textquotedblleft discriminatory\textquotedblright%
\ facilities that could alternatively have been invoked in the game
representations. All proofs (based entirely on elementary linear algebra) are
in Section \ref{proofs} at the end. Finally, the numbering system used in this
paper (and induced by Scientific Workplace) stems from the arrangement of all
definitions, theorems, etc. in \emph{one} grand sequence (rather than separate
sequences for each category). Thus we have Definitions 1,....5, followed by
Remark 6, then Definitions 7,8, then Theorem 9 (the first theorem of the
paper), and so on, making it hopefully more convenient to locate them.

\section{Games\label{games}}

Throughout (unless otherwise stated) \emph{all} sets\ are assumed to be finite
and non-empty.

Let $N$ denote a set of \emph{players, }$\mathcal{C\subseteq N=}\left\{
S:\phi\neq S\subseteq N\right\}  $ a collection of \emph{coalitions}, and
$\mathbb{R}^{\mathcal{C}}$ the Euclidean space whose axes are indexed by the
coalitions in $\mathcal{C}$. We assume that $N$ is the union of all the
coalitions in $\mathcal{C}$:
\begin{equation}
N=\cup\left\{  S:S\in\mathcal{C}\right\}  \label{=all players}%
\end{equation}
i.e., in our model, a player exists only via membership in some coalition
(possibly, just the singleton coalition consisting of himself).

A \emph{game}\footnote{The empty set $\emptyset$ is customarily admitted in
$\mathcal{C}$ with the stipulation that $v\left(  \emptyset\right)  =0.$ We
exclude $\emptyset$ from $\mathcal{C}$ for notational convenience.} on
$\mathcal{C}$ is a vector $v\in\mathbb{R}^{\mathcal{C}}$ (equivalently, a
function $v:\mathcal{C\longrightarrow}$ $\mathbb{R}$).

\section{Assignments \label{assignments}}

There is a set $K$ of \emph{facilities}. Let $\mathcal{K}$ denote the
collection of subsets of $K$. The map $\psi:N\longrightarrow\mathcal{K}$
assigns a set $\psi(n)\in\mathcal{K}$ of facilities that player $n\in N$ must
use. For each facility $k\in K$, its \emph{full user-set} is $\psi^{-1}\left(
k\right)  =\left\{  n\in N:k\in\psi(n)\right\}  $ and assumed to be non-empty.
Conditional on the coalition $S\in\mathcal{C}$, the user-set at $k$ is
$\psi^{-1}\left(  k\right)  \cap S$ (which could be empty).

There is another map $\gamma:N\longrightarrow\mathbb{R}$ which assigns
\emph{cost}\textit{ }$\gamma(k)\in\mathbb{R}$ to facility $k$, that accrues
collectively to the users of $k,$ regardless of who or how many they
are.\footnote{One might think of $\gamma\left(  k\right)  $ as a fixed cost
(or, subsidy) that accompanies the establishment of facility $k.$}

The pair of maps $(\psi,\gamma)$ is called an \emph{assignment}.

In the sequel, $N$ will be fixed, while the maps $\psi,\gamma$ are allowed to
vary in order to generate different assignments\footnote{This of course
entails varying $K$.}. (One may visualize $(\psi,\gamma)$ as a bipartite
graph, with disjoint sets $N,K$; an edge $\left(  n,k\right)  $ if, and only
if, $k\in\psi(n);$ and cost $\gamma(k)$ at each node $k\in K.$ 

Given the assignment $(\psi,\gamma),$ the natural question arises: how
\textquotedblleft should\textquotedblright\ the total cost $\sum_{k\in
K}\gamma(k)$ be allocated among the users in $N$?

\begin{definition}
The \emph{equitable solution} to $\left(  \psi,\gamma\right)  $ is given by
$\tau(\psi,\gamma)=\left(  \tau_{n}(\psi,\gamma)\right)  _{n\in N}$ where%
\begin{equation}
\tau_{n}(\psi,\gamma)=\sum_{k\in\psi(n)}\frac{\gamma(k)}{\left\vert \psi
^{-1}\left(  k\right)  \right\vert } \label{equitable soln}%
\end{equation}
(Since $\left\vert \psi^{-1}\left(  k\right)  \right\vert >0$ for every $k\in
K,$ the vector $\tau$ is well-defined.)
\end{definition}

\section{Game Representations\label{representations}}

\begin{definition}
\label{representation def}The assignment $\left(  \psi,\gamma\right)  $ is a
\emph{representation} of the game $v\in\mathbb{R}^{\mathcal{C}}$ if, denoting
$\psi(S)=\cup_{n\in S}\psi\left(  n\right)  ,$ we have
\begin{equation}
v(S)=\sum_{k\in\psi(S)}\gamma(k)\text{ for all }S\in\mathcal{C}
\label{Game Rep 1}%
\end{equation}
and
\begin{equation}
\psi^{-1}\left(  k\right)  \in\mathcal{C}\text{ for all }k\in K.
\label{Game Rep 2}%
\end{equation}

\end{definition}

Condition (\ref{Game Rep 1}) simply says that $v\left(  S\right)  $ is the
total cost of the facilities in $\psi(S)$ that players in $S$ are assigned to
use, for every coalition $S$ in $\mathcal{C}$ (see Section
\ref{superadditive subsection} for the game-theoretic model behind
(\ref{Game Rep 1})). Condition (\ref{Game Rep 2}) amounts to the measurability
of $\psi$ w.r.t. $\mathcal{C}$, and is a natural consistency requirement for
$\left(  \psi,\gamma\right)  $ to represent a game with underlying coalitions
$\mathcal{C}$. (It is also necessary for our analysis, see Comment 2 in
Section \ref{comments}.)

Given any representation $\left(  \psi,\gamma\right)  $ of $v\in
\mathbb{R}^{\mathcal{C}},$ there exist infinitely many other representations
of $v$ that arise trivially from $\left(  \psi,\gamma\right)  $ by adding
facilities with zero cost or by splitting facilities.

\begin{definition}
We say that $\left(  \psi^{\prime},\gamma^{\prime}\right)  $ is a
\emph{trivial expansion} of $\left(  \psi,\gamma\right)  $ if it is obtained
from $\left(  \psi,\gamma\right)  $ via the following two steps (where either
step could be skipped). \textbf{Step 1}: add finitely many facilities
$k^{\prime}$ to those of $\left(  \psi,\gamma\right)  ,$ with $\gamma\left(
k^{\prime}\right)  =0$ and the full user-set of $k^{\prime}$ chosen
arbitrarily from $\mathcal{C}.$ \textbf{Step 2:} Replace each facility
$k^{\prime\prime}$ of the enlarged assignment created in Step 1, with any
finite number of replicas, setting arbitrary (negative or positive) costs for
each replica, subject to the constraint that (i) the total cost of all the
replicas of $k^{\prime\prime}$ add up to $\gamma\left(  k^{\prime\prime
}\right)  ;$ and (ii) the full user-set of every replica of $k^{\prime\prime}$
is the same as the full user-set of $k^{\prime\prime}.$
\end{definition}

It is evident that (i) if $\left(  \psi,\gamma\right)  $ is a representation
of $v\in\mathbb{R}^{\mathcal{C}}$ then so are all its trivial expansions; and
(ii) all trivial expansions of $\left(  \psi,\gamma\right)  $ have the same
equitable solution. We shall show that, but for the innocuous multiplicity
created by trivial expansions, each game $v$ has a \emph{unique}
representation (see Theorem \ref{general TU} in Section
\ref{Ext Shapley Value}).

\section{The Shapley Value\label{Shapley Value}}

In the \textquotedblleft canonical\textquotedblright\ case $\mathcal{C=N}$,
the \emph{Shapley value} of the game $v\in\mathbb{R}^{\mathcal{N}}$ is the
vector $\varphi(v)=\left(  \varphi_{n}\left(  v\right)  \right)  _{n\in N}%
\in\mathbb{R}^{N}$ where (admitting the empty set $\emptyset\subseteq
N\diagdown\left\{  n\right\}  $ in the formula below): $:$%
\[
\varphi_{n}\left(  v\right)  =\sum_{S\subseteq N\diagdown\left\{  n\right\}
}\frac{\left\vert S\right\vert !(\left\vert N\right\vert -1-\left\vert
S\right\vert )!}{\left\vert N\right\vert !}\left[  v\left(  S\cup\left\{
n\right\}  \right)  -v(S)\right]
\]
Here $\left\vert X\right\vert $ denotes the cardinality of the set $X$, and
$v\left(  \phi\right)  $ is understood to be $0,$ and $0!$ to be $1.$ As shown
in (\cite{Shapley 1951}, \cite{Shapley 1953}), $\varphi:\mathbb{R}%
^{\mathcal{N}}\longrightarrow\mathbb{R}^{N}$ is the unique map that satisfies
certain intuitively appealing axioms named \textquotedblleft
dummy\textquotedblright, \textquotedblleft symmetry\textquotedblright%
,\textquotedblleft additivity\textquotedblright\ and \textquotedblleft
efficiency\textquotedblright, the last tantamount to $\sum_{n\in N}\varphi
_{n}\left(  v\right)  =v\left(  N\right)  ,$ implying that $\varphi\left(
v\right)  $ is an allocation of the total proceeds $v\left(  N\right)  $ in
the game$.$

Note that the formula is \emph{not} defined for $v\in\mathbb{R}^{\mathcal{C}}$
when $\mathcal{C\subsetneqq N}$, since there exist $S\in\mathcal{C}$ such that
$S$ or $S\cup\left\{  n\right\}  $ fail to be in $\mathcal{C}.$

\section{Conditions on Coalitions\label{Curtailed}}

\begin{definition}
\label{MM games}($MM$-\textbf{Games}) For any two coalitions $S$ and $T$, we
say that $S$ \emph{meets }$T$ if $S\cap T\neq\phi$; and that $S$ \emph{misses
}$T$ if $S\cap T=\phi.$ Given $\mathcal{C\subseteq N}$ and $S\in\mathcal{C},$
the $MM$\textbf{-}\emph{game} $w_{S}=\left(  w_{S}(T)\right)  _{T\in
\mathcal{C}}\in\mathbb{R}^{\mathcal{C}}$ is defined by

$w_{S}(T)=\left\{
\begin{array}
[c]{ccc}%
1 & \text{if} & S\text{ meets }T\\
0 & \text{if} & S\text{ misses }T
\end{array}
\right.  $
\end{definition}

\begin{definition}
(\textbf{Full Span}) $\mathcal{C}$ has \emph{full span }if the vectors
$\left\{  w_{S}\right\}  _{S\in\mathcal{C}}$ form a basis of $\mathbb{R}%
^{\mathcal{C}}.$ (In this case $\left\{  w_{S}\right\}  _{S\in\mathcal{C}}$ is
called the $MM$\emph{-basis }of $\mathbb{R}^{\mathcal{C}}.$)
\end{definition}

\begin{remark}
\label{matrix}(\textbf{Matrix Version).}Consider the square $\left(
0,1\right)  $-matrix $A$ whose rows and columns are indexed by $\mathcal{C}$,
and whose column vectors are $w_{S}$ for $S\in\mathcal{C}$. Then $\mathcal{C}$
has full span $\Longleftrightarrow$ $A$ is invertible.
\end{remark}

The full span condition is exactly what is needed for our main results (see
Theorems \ref{general TU} and \ref{necessity of full span} below). However the
condition is \textquotedblleft indirect\textquotedblright, and it may be
operationally useful to formulate some purely set-theoretic (albeit more
stringent) variants. We present two such below.

\begin{definition}
(\textbf{Hierarchy) }$\mathcal{C}$ is \emph{hierarchical} if the elements of
$\mathcal{C}$ can be arranged in a sequence $S_{1},\ldots,S_{l}$ such that,
for $k=2,\ldots,l,$%
\[
\left(  i\right)  \text{ }S_{1}\text{ meets }S_{k}\text{ }%
\]
and
\[
\left(  ii\right)  \text{ there exists }T\in\mathcal{C}\text{ which misses
}S_{k}\text{ and meets }S_{1},\ldots,S_{k-1}%
\]
In this case, the sequence $S_{1},\ldots,S_{l}$ is called a \emph{hierarchy}
for $\mathcal{C}$.
\end{definition}

By way of interpretation, think of \textquotedblleft$S$ misses $T$%
\textquotedblright\ as \textquotedblleft$S$ rejects $T$\textquotedblright, and
\textquotedblleft$S$ meets $T$\textquotedblright\ as \textquotedblleft$S$
accepts $T$\textquotedblright. Then the idea behind a hierarchy is that it
should be possible to reject coalitions in $\mathcal{C}$ sequentially until a
sole, final survivor ($S_{1}$ in our notation) is left$.$ At any stage $k$ of
the sequence (starting at $k=l$ and ending at $k=2$), some coalition
$T\in\mathcal{C}$ must be at hand to reject one of the coalitions that have
survived until that stage, but (in order to give $T$ full sanction to do the
rejecting) all the other survivors must \textquotedblleft
accept\textquotedblright\ $T$.

Hierarchies for $\mathcal{C}$, if they exist, need not be unique though
clearly they all have the same first element (final survivor). It is easy to
construct examples of $\mathcal{C}$ with several different hierarchies (indeed
we shall do so in the second half of the proof of Theorem \ref{Lemma} in
Section \ref{proofs}, and also in Section \ref{Examples}).

\begin{definition}
\textbf{(Semi-Algebra) }$\mathcal{C}$ is a \emph{semi-algebra}\footnote{In
other words, $\mathcal{C}$ is a semi-algebra if $\mathcal{C\cup}\left\{
\emptyset\right\}  $ is closed under complementation. Recall that
$\mathcal{C}$ is called an \emph{algebra} if, in addition to (i) and (ii),
$\mathcal{C}$ is closed under unions and (non-empty) intersections. (In the
standard definition, the empty set is included in $\mathcal{C}$, and so the
clause \textquotedblleft non-empty\textquotedblright\ is dropped.) Since $N$
is finite, any algebra on $N$ is generated by a partition of $N$ (by taking
unions). The union of finitely many algebras is a semi-algebra on $N$ but not
an algebra.}\emph{ }if\emph{ }(i) $N\in\mathcal{C}$; and (ii) $S\in
\mathcal{C\diagdown}\left\{  N\right\}  \Longrightarrow N\diagdown
S\in\mathcal{C}.$
\end{definition}

\begin{theorem}
\label{Lemma}$\mathcal{C}$ is a semi-algebra $\Longrightarrow$ $\mathcal{C}$
is hierarchical $\Longrightarrow$ $\mathcal{C}$ has full span.
\end{theorem}

\begin{proof}
See Section \ref{proofs}.
\end{proof}

\section{Extensions of the Shapley Value\label{Ext Shapley Value}}

\begin{theorem}
(\textbf{Existence)}\label{general TU} Assume $\mathcal{C}$ has full span, and
let $v\in\mathbb{R}^{\mathcal{C}}.$ There exists a unique representation
$\left(  \psi^{\ast},\gamma^{\ast}\right)  $ of $v$ with the minimum number of
facilities. All other representations of $v$ are trivial expansions of
$\left(  \psi^{\ast},\gamma^{\ast}\right)  $. Thus the equitable solution is
invariant across all representations of $v$. Finally, in the canonical case
$\mathcal{C=N}$, the equitable solution is the Shapley value of $v$.
\end{theorem}

\begin{proof}
See Section \ref{proofs}.
\end{proof}

\begin{definition}
Assume $\mathcal{C}$ has full span. The \emph{equitable solution }on games in
$\mathbb{R}^{\mathcal{C}}$ is the map
\[
\chi:\mathbb{R}^{\mathcal{C}}\longrightarrow\mathbb{R}^{N}%
\]
given by%
\[
\chi\left(  v\right)  =\tau\left(  (\psi,\gamma)\right)  \text{ where }%
(\psi,\gamma)\text{ is any representation of }v
\]

\end{definition}

In view of Theorem \ref{general TU}, all representations $(\psi,\gamma)$ of
$v$ have the same equitable solution $\tau\left(  (\psi,\gamma)\right)  $, so
the definition of $\chi$ makes sense.

We next highlight some properties of the extension $\chi.$ It is natural to
consider two players to be \textquotedblleft symmetric\textquotedblright\ in
the game $v:\mathcal{C\longrightarrow}$ $\mathbb{R}$ if their roles are
interchangeable in $v$, i.e., replacing any one of them by the other in a
coalition, alters neither\emph{ the coalition's membership in }$\mathcal{C}$
nor its payoff.\footnote{The membership (resp., payoff) requirement reflects
symmetry in the domain (resp., range) of $v$. In the canonical case
$\mathcal{C=N}$, the membership requirement holds automatically, and our
definition reduces to the standard one.} Formally

\begin{definition}
\label{defn symmetry}Players $i$ and $j$ are \emph{symmetric }in the game
$v\in\mathbb{R}^{\mathcal{C}}$ if, for all $T\in\mathcal{N}\mathcal{\cup
}\left\{  \emptyset\right\}  $ such that $i\notin T$ and $j\notin T,$ we have
\[
T\cup\left\{  i\right\}  \in\mathcal{C}\Longleftrightarrow T\cup\left\{
j\right\}  \in\mathcal{C}%
\]
and%
\[
v\left(  T\cup\left\{  i\right\}  \right)  =v\left(  T\cup\left\{  j\right\}
\right)  \text{ if }T\cup\left\{  i\right\}  \in\mathcal{C}%
\]

\end{definition}

In the same vein, we define a player to be a \textquotedblleft
dummy\textquotedblright\ if adding him to a coalition is irrelevant for both
the coalition's membership in $\mathcal{C}$ and its payoff, i.e.,

\begin{definition}
\label{defn dummy}Player $i$ is a \emph{dummy} in $v$ if (i) $v\left(
i\right)  =0$ when $\left\{  i\right\}  \in\mathcal{C}$; and (ii) for all
$T\in\mathcal{N}$ such that $i\notin T$:
\[
T\cup\left\{  i\right\}  \in\mathcal{C}\iff T\in\mathcal{C}%
\]
and%
\[
v\left(  T\cup\left\{  i\right\}  \right)  =v\left(  T\right)  \text{ if }%
T\in\mathcal{C}%
\]

\end{definition}

\begin{theorem}
(\textbf{Properties)}\label{properties} Assume $\mathcal{C}$ has full span.
Then the extension $\chi:\mathbb{R}^{\mathcal{C}}\longrightarrow\mathbb{R}%
^{N}$ has the following three properties: (Symmetry) if $i$ and $j$ are
symmetric in $v,$ then $\chi_{i}\left(  v\right)  =\chi_{j}\left(  v\right)
;$ (Dummy) if $i$ is a dummy in $v,$ then $\chi_{i}\left(  v\right)  =0$;
(Linearity)$\chi\left(  v+w\right)  =\chi\left(  v\right)  +\chi\left(
w\right)  $ and $\chi\left(  cv\right)  =c\chi\left(  v\right)  $ for any
$v,w$ and scalar $c.$
\end{theorem}

\begin{proof}
See Section \ref{proofs}.
\end{proof}

\begin{remark}
Define a $\mathcal{C}$\emph{-assignment} to be a pair $(\psi,\gamma)$, where
the map $\psi:N\longrightarrow\mathcal{K}$ satisfies condition
(\ref{Game Rep 2}), i.e., $\psi^{-1}\left(  k\right)  \in\mathcal{C}$ for all
$k\in K;$ and $\gamma:K\longrightarrow\mathbb{R}$ is arbitrary. Let
$\mathfrak{C}$ denote the collection of all $\mathcal{C}$-assignments for any
fixed $\mathcal{C}$; and for every $(\psi,\gamma)\in\mathfrak{C,}$ let
$\zeta\left(  (\psi,\gamma)\right)  $ denote the unique game in $\mathbb{R}%
^{\mathcal{C}}$ which is represented by $(\psi,\gamma)$. Then Theorem
\ref{general TU} may be restated as follows: if $\mathcal{C}$ has full span,
the map $\zeta:\mathfrak{C\longrightarrow}\mathbb{R}^{\mathcal{C}}$ is onto
and, for any $v\in\mathbb{R}^{\mathcal{C}},$ the fiber $\zeta^{-1}\left(
v\right)  $ consists of all representations of $v$; moreover $\tau\left(
(\psi,\gamma)\right)  =\chi\left(  v\right)  $ for all $(\psi,\gamma)\in
\zeta^{-1}\left(  v\right)  .$
\end{remark}

Finally, let us point out that the condition of full span on $\mathcal{C}$ is
not only sufficient, but also necessary, for the conclusion of Theorem
\ref{general TU} (and thus cannot be weakened). To be precise, define%
\[
Sp\left(  \mathcal{C}\right)  =\text{Span}\left\{  w_{S}:S\in\mathcal{C}%
\right\}  \subset\mathbb{R}^{\mathcal{C}}%
\]
where (recall) $w_{S}$ is the $MM$ game of Definition \ref{MM games}. Thus
$\mathcal{C}$ has full span if, and only if, $Sp\left(  \mathcal{C}\right)
=\mathbb{R}^{\mathcal{C}}$.

\begin{theorem}
(\textbf{Necessity of Full Span)}\label{necessity of full span} (i)
$v\in\mathbb{R}^{\mathcal{C}}$ has a representation $\Longleftrightarrow$
$v\in Sp\left(  \mathcal{C}\right)  $ (and thus: every $v\in\mathbb{R}%
^{\mathcal{C}}$ has a representation $\Longleftrightarrow$ $\mathcal{C}$ has
full span). (ii) For every $v\in Sp\left(  \mathcal{C}\right)  ,$ all
representations of $v$ have the same equitable solution $\Longleftrightarrow$
$\mathcal{C}$ has full span.
\end{theorem}

\begin{proof}
See Section \ref{proofs}.
\end{proof}

\section{Games on Curtailed Coalitions: some Examples\label{Examples}}

\subsection{Example 1 (Semi-Algebra)}

Consider the following scenario. Each $n\in N$ specializes in producing a
distinctive good (also labeled $n$ for convenience) and sends it on to a
retailer for sale. The retailer, having received all the goods $n\in N,$ can
partition them into \textquotedblleft bundles\textquotedblright\ (subsets of
$N$) prior to sale, though\emph{ not all partitions may be
feasible\footnote{For example, a multi-national retailer will not combine
left-hand drive and right-hand-drive cars within a bundle meant for a national
market. It will not even experiment with such incongruous bundles to estimate
how much financial loss they might cause. So instead of imputing some
arbitrary negative value to such bundles, it makes more sense to deem them
infeasible.}}. Let $\mathcal{P}_{1},\ldots,\mathcal{P}_{k}$ denote the
collection of feasible partitions of $N$. Now the revenue $v_{i}\left(
S\right)  =v\left(  S,\mathcal{P}_{i}\right)  $ earned by a bundle
$S\in\mathcal{P}_{i}$ may well depend upon the partition $\mathcal{P}_{i}$ to
which $S$ belongs\footnote{e.g., on account of \textquotedblleft
externalities\textquotedblright: two bundles may be complements and boost each
other's sales, or they may be substitutes and diminish the sales.}. The
retailer selects a partition which maximizes total sales revenue. The question
arises: how \textquotedblleft should\textquotedblright\ this total be split
amongst the producers $n\in N$?

To this end, our extension of the Shapley value in Section
\ref{Ext Shapley Value} is of relevance. The above scenario can be modeled as
a game on a \emph{semi-algebra} of coalitions, that arises from auxiliary
games in \textquotedblleft partition function form\textquotedblright.

Consider finitely many \textquotedblleft auxiliary\textquotedblright\ games
$v_{1},\ldots,v_{k}$ defined on algebras $\mathcal{C}_{1},\ldots
,\mathcal{C}_{k}$ whose atoms are given by the feasible partitions
$\mathcal{P}_{1},\ldots,\mathcal{P}_{k}$., i.e., $\mathcal{C}_{i}$ is the
collection of all sets generated by taking unions of the elements of
$\mathcal{P}_{i}$. The games $v_{i}:\mathcal{C}_{i}\longrightarrow\mathbb{R}$
are defined as follows: if $T=\cup_{j}S_{j}$ for $S_{j}\in\mathcal{P}_{i}$,
then%
\begin{equation}
v_{i}\left(  T\right)  =\sum_{j}v_{i}\left(  S_{j}\right)
\label{superadditivity}%
\end{equation}
Now the union $\mathcal{C=\cup}_{i=1}^{k}\mathcal{C}_{i}$ of the algebras is
\emph{not} an algebra in general but a \emph{semi-algebra}, consisting of all
the coalitions that are compatible with one of the permitted partitions
$\mathcal{P}_{1},\ldots,\mathcal{P}_{k}$. The game $v$ on $\mathcal{C}$ arises
naturally as follows\footnote{Had we supposed $v_{i}$ to be given in a more
general form, with a \textquotedblleft superadditivity\textquotedblright%
\ property (that replaces \textquotedblleft$=$\textquotedblright\ with
\textquotedblleft$\geq$\textquotedblright\ in (\ref{superadditivity}) ), then
again $v$ could be defined from the $v_{i}$ as before. In the absence of
superadditivity, however, the definition of $v$ can become more subtle than in
(\ref{partition function}). Any coalition $S$ will now consider various
possibilities of partitioning itself that are consistent with one of
$\mathcal{P}_{1},\ldots,\mathcal{P}_{k}.$ However, given any such partition of
$S$, the players in $N\diagdown S$ may in respond by a partition of their own
(so long as the two partitions of $S$ and $N\diagdown S$ are compatible with
one of $\mathcal{P}_{1},\ldots,\mathcal{P}_{k}$) in order to \emph{minimize}
the payoff to $S.$ Being cautious, $S$ should hypothesize that this will
indeed happen; and, under this hypothesis, $S$ should partition itself in the
best possible way. This will imply a $\max\min$ formula for $v\left(
S\right)  .$
\par
Other methods for defining $v\left(  S\right)  $ could also be thought of,
which entail less pessimism regarding the expectations of $S$. However, no
matter how we define $v\left(  S\right)  $, our point here remains intact,
namely: the relevant game $v$ is defined on a \emph{semi-algebra }of
coalitions, not on all of $\mathcal{N}$.}:%
\begin{equation}
v\left(  S\right)  =\min\left\{  v_{i}\left(  S\right)  :S\in\mathcal{C}%
_{i}\text{ for some }1\leq i\leq k\right\}  \label{partition function}%
\end{equation}
The see the rationale for (\ref{partition function}), recall that $v\left(
S\right)  $ is meant to be the payoff that the coalition $S$ of producers can
\emph{guarantee} to itself regardless of what the others in $N\diagdown S$ do.
Since $S$ cannot be sure of which $\mathcal{C}_{i}$ it could end up in (as
this depends on the configuration of the others), it must look at the
worst-case scenario, which accounts for the $\min$ term in
(\ref{partition function}).

\subsection{Example 2 (Hierarchy)}

Consider players/parties $a,b,c,d,e$ arranged from left to right in a
social/political spectrum. The extremes $a,e$ can only get along with their
immediate neighbors $b,d$ respectively. However $b$ and $d$ are able to extend
their hand \textquotedblleft across the aisle\textquotedblright\ to each other
provided $c$ is present to intermediate between them. As for $c,$ he can only
function as an intermediary. Assuming, finally, that at least two persons are
needed for a coalition to be viable, we get that $\mathcal{C=}\left\{
\left\{  a,b\right\}  ,\left\{  b,c,d\right\}  ,\left\{  d,e\right\}
\right\}  $. Clearly $\mathcal{C}$ is hierarchical, with two possible
hierarchies $\left\{  b,c,d\right\}  ,\left\{  a,b\right\}  ,\left\{
d,e\right\}  $ and $\left\{  b,c,d\right\}  ,\left\{  d,e\right\}  ,\left\{
a,b\right\}  .$ Thus Theorem \ref{general TU} applies and the extended Shapley
value $\chi$ is well-defined for all games $v:\mathcal{C\longrightarrow
}\mathbb{R}$. By way of illustration, consider the cost-allocation game given
by $v(a,b)=-18,$ $v\left(  d,e\right)  =-16,$ $v(b,c,d)=-22.$ Here $\chi$
allocates costs to the players as follows: $\chi_{a}\left(  v\right)  =-3,$
$\chi_{b}\left(  v\right)  =-7,$ $\chi_{c}\left(  v\right)  =-4,$ $\chi
_{d}\left(  v\right)  =-6,$ $\chi_{e}\left(  v\right)  =-2$. This can be seen
from the unique minimal representatation $\left(  \psi,\gamma\right)  $ of
$v,$ given as follows. There are three facilities $k_{1},k_{2,}k_{3}$
corresponding to $\left\{  a,b\right\}  ,\left\{  b,c,d\right\}  ,\left\{
d,e\right\}  .$ The correspondence $\psi$ is given by $\psi\left(  a\right)
=\left\{  k_{1}\right\}  ,\psi\left(  b\right)  =\left\{  k_{1},k_{2}\right\}
,\psi\left(  c\right)  =\left\{  k_{2}\right\}  ,\psi\left(  d\right)
=\left\{  k_{2},k_{3}\right\}  ,\psi\left(  e\right)  =\left\{  k_{3}\right\}
.$ The map $\gamma$ is given by $\gamma\left(  k_{1}\right)  =-6,$
$\gamma\left(  k_{2}\right)  =-12,$ $\gamma\left(  k_{3}\right)  =-4$.

Note that, without going to the representation $\left(  \psi,\lambda\right)
$, not only is the cost allocation $\chi\left(  v\right)  $ unknown, it is not
even clear what the total cost is that needs to be allocated (since there is
no $v\left(  N\right)  $ given upfront). Only when we represent\ $v$ by
$\left(  \psi,\lambda\right)  $, does $\chi\left(  v\right)  $ comes to light.

There is an algorithm to extend $v:\mathcal{C\longrightarrow}\mathbb{R}$ to a
game $\widetilde{v}:\mathcal{N\longrightarrow}\mathbb{R}$ on the canonical
domain $\mathcal{N}$ of all coalitions, such that $\chi\left(  v\right)  $ is
the Shapley value of $\widetilde{v}$. But the algorithm is not obvious and
entails assigning numerical values $v\left(  S\right)  $ for $S\in
\mathcal{N\diagdown C}$ in a very particular way (see Comment 5 in Section
\ref{comments}). For instance, were we to assign $0$ to all the coalitions in
$\mathcal{N\diagdown C}$ (which might be one's first natural instinct), the
total allocated by the Shapley value to the five players would be
$\widetilde{v}\left(  N\right)  =0$ which is a far cry from $\chi\left(
v\right)  .$

\section{Discriminatory Facilities?\label{Discriminatory Facilities}}

\subsection{From Assignments to Games\label{assignments to games}}

In our representation $\left(  \psi,\gamma\right)  $ of the game
$v\in\mathbb{R}^{\mathcal{C}},$ the map $\psi$ can be discriminatory,
assigning different sets $\psi(n)\subseteq K$ to players $n\in N$. However
recall that each facility $k\in K$ is \emph{user-blind} in two senses of that
word. First, $k$ is \emph{decentralized }in that $\gamma(k)$ is independent of
the user-sets, or the costs incurred, at the other facilities in
$K\diagdown\left\{  k\right\}  .$ Second, $k$ is \emph{non-discriminatory }in
that $\gamma(k)$ is\textit{ }even independent of the user-set at $k.$

Indeed, like Themis (the goddess of justice), the facility $k$ in our model
wears a true blindfold towards all users, both at its own site and elsewhere,
not even keeping track of how many there are, leave aside their identities.
And, \emph{once we position ourselves at facility }$k$\emph{ and adopt the
}$k$\emph{-standpoint, it is precisely this blindfold that provides a
compelling rationale for allocating }$\gamma(k)$\emph{ equally among the
players in }$\psi^{-1}\left(  k\right)  $\emph{ at each }$k\in K.$

One could think of other game representations, based on facilities that are
\emph{discriminatory} (though, for simplicity, we shall continue to assume
that they are decentralized). Then this rationale breaks down and it is no
longer clear how to allocate $\gamma(k)$ among $\psi^{-1}\left(  k\right)  $,
as we shall now see.

In order to incorporate discriminatory facilities in our representations, we
need to first enhance the definition of $\gamma$. Consider a map
$\psi:N\longrightarrow\mathcal{K}$ as before, which assigns facilities
$\psi\left(  n\right)  \subset K$ that each $n\in N$ must use; and a
collection $\mathcal{C}$ of feasible coalitions of $N$. Any $S\in\mathcal{C}$
may come to the representation $\left(  \psi,\gamma\right)  $ to determine its
worth $v_{\left(  \psi,\gamma\right)  }\left(  S\right)  $ in $\left(
\psi,\gamma\right)  $. Conditional on $S\in\mathcal{C},$ the user-set sent by
$\psi$ to facility $k\in K$ is $\psi^{-1}\left(  k\right)  \cap S$. Thus the
collection of potential user-sets at $k$ is
\begin{equation}
\mathcal{K}_{k}=\mathcal{K}_{k}\left(  \psi,\mathcal{C}\right)  =\left\{
\psi^{-1}\left(  k\right)  \cap S:S\in\mathcal{C}\right\}
\label{potential users at  k}%
\end{equation}
(Note that many sets in $\mathcal{K}_{k}$ may \emph{not} be in\footnote{For we
are not assuming that $\mathcal{C}$ is closed under intersections.}
$\mathcal{C}$ and, in particular, one such could be the empty set $\emptyset
$.) The assignment of costs to user-sets at $k$ comprises a map\footnote{We
use $\gamma_{k}$ to denote the map, even though it denoted a scalar earlier.
This should cause no confusion, as the meaning will always be clear from the
context.}
\begin{equation}
\gamma_{k}:\mathcal{K}_{k}\longrightarrow\mathbb{R} \label{computer program}%
\end{equation}
(with the usual convention $\gamma_{k}\left(  \emptyset\right)  =0$). In this
set-up, $k$ can be considered user-blind \emph{only when} $\gamma_{k}$ is a
constant map, for then $\mathcal{K}_{k}$ has no role to play. Indeed the
facility $k$ may simply post the constant cost $\gamma_{k}\in\mathbb{R}$,
turning its back upon the user-sets $\mathcal{K}_{k}$ sent to it by $\psi$.
However, if $\gamma_{k}\left(  T\right)  $ varies at all with $T$, it behoves
$k$ to scrutinize every user-set $T\in\mathcal{K}_{k}$ in order to assign the
costs $\gamma_{k}\left(  T\right)  $. This leads us to define $k$ to be a
\emph{discriminatory facility} if, and only if, the map $\gamma_{k}$ is not a constant.

Denoting, without confusion, the family of maps $\left\{  \gamma_{k}\right\}
_{k\in K}$ also by $\gamma,$ the assignment $\left(  \psi,\gamma\right)  $
allows for discriminatory facilities.

\begin{definition}
\label{defn induced game}The game $v_{\left(  \psi,\gamma\right)  }\in
R^{\mathcal{C}}$ induced by $\left(  \psi,\gamma\right)  $ is given by
\begin{equation}
v_{\left(  \psi,\gamma\right)  }\left(  S\right)  =\sum_{k\in\psi\left(
S\right)  }\gamma_{k}\left(  \psi^{-1}\left(  k\right)  \cap S\right)  \text{
for }S\in\mathcal{C}\label{formula for v}%
\end{equation}

\end{definition}

\begin{definition}
\label{defn local game}The local game $v_{\left(  \psi,\gamma\right)  }^{k}\in
R^{\mathcal{C}}$ induced by $\left(  \psi,\gamma\right)  $ at facility $k$ is
given by
\begin{equation}
v_{\left(  \psi,\gamma\right)  }^{k}\left(  S\right)  =\gamma_{k}\left(
\psi^{-1}\left(  k\right)  \cap S\right)  \text{ for }S\in\mathcal{C}%
\label{formula for local game}%
\end{equation}

\end{definition}

Clearly
\begin{equation}
v_{\left(  \psi,\gamma\right)  }=%
{\textstyle\sum_{k\in K}}
v_{\left(  \psi,\gamma\right)  }^{k} \label{sum of local games}%
\end{equation}

\begin{definition}
\label{Representation with Discriminatory Facilities} The assignment $\left(
\psi,\gamma\right)  $ represents the game $v\in\mathbb{R}^{\mathcal{C}}$ if%
\begin{align*}
v_{\left(  \psi,\gamma\right)  }  &  =v\\
\psi^{-1}\left(  k\right)   &  \in\mathcal{C}\text{ for all }k\in K
\end{align*}

\end{definition}

Note that Definition \ref{Representation with Discriminatory Facilities}
generalizes our formulae (\ref{Game Rep 1}), (\ref{Game Rep 2}) given earlier
for the special case of user-blind facilities where the map $\gamma_{k}$ was a
constant (i.e, $\gamma_{k}\left(  T\right)  =\gamma_{k}\in\mathbb{R}$ for all
$T\in\mathcal{K}_{k}$).

\subsection{Superadditive Games: A Rationale for Definition
\ref{Representation with Discriminatory Facilities}%
\label{superadditive subsection}}

From the game-theoretic point of view, Definition \ref{formula for v} (and
thus Definition \ref{Representation with Discriminatory Facilities}) is best
justified in the context of superadditive games (defined below), even though
our \emph{formal} analysis does not depend on superadditivity.

\begin{definition}
\label{superadditivity defn}For any $S\in\mathcal{C}$, a partition
$S=S_{1}\cup\ldots\cup S_{l}$ of $S$ is $\mathcal{C}$-\emph{feasible} if every
$S_{i}\in\mathcal{C}.$ A game $v\in\mathbb{R}^{\mathcal{C}}$ is
\emph{superadditive\footnote{In the canonical case $\mathcal{C=N}$, Definition
\ref{superadditivity defn} corresponds to the standard definition, namely:
$v\left(  S\cup T\right)  \geq v\left(  S\right)  +v\left(  T\right)  $
whenever $S\cap T\neq\emptyset.$} }if
\begin{equation}
v\left(  S\right)  \geq v\left(  S_{1}\right)  +\ldots+v\left(  S_{l}\right)
\end{equation}
for every $\mathcal{C}$-feasible partition of $S.$
\end{definition}

We now make explicit a (so-far tacit) game-theoretic model for the role of
coalitions in a representation $\left(  \psi,\gamma\right)  $, which directly
leads to Definition \ref{Representation with Discriminatory Facilities}. First
recall that each player $n$ \emph{must }use facilities $\psi\left(  n\right)
\subset K$. For this to make sense, we need to make the concomitant postulate
that $n$ should be able to access (or, set up) $\psi\left(  n\right)  $,
without interference from other players. It follows that a coalition
$S\in\mathcal{C}$ can obtain the payoff $\sum_{k\in\psi(S)}\gamma_{k}\left(
\psi^{-1}\left(  k\right)  \cap S\right)  $ from the facilities $\psi(S)$ that
it has access to (without interference from players in $N\diagdown S$)$.$ It
is natural to furthermore postulate that $S$ has the strategic freedom to
create any $\mathcal{C}$-feasible partition $S_{1},\ldots,S_{l}$ of itself
before coming to $\left(  \psi,\gamma\right)  $, in which case it can obtain
payoff $\sum_{k\in\psi(S_{i})}\gamma_{k}\left(  \psi^{-1}\left(  k\right)
\cap S_{i}\right)  $ in $\left(  \psi,\gamma\right)  $ via its subcoalition
$S_{i}$ for $1\leq i\leq l$. Thus $S$ can in fact obtain $\max$ $\sum
_{k\in\psi(S_{1})}\gamma\left(  k\right)  +\ldots+\sum_{k\in\psi(S_{l})}%
\gamma\left(  k\right)  $ where the $\max$ is taken over all $\mathcal{C}%
$-feasible partitions of $S$. If $v$ is superadditive the $\max$ is
$\sum_{k\in\psi\left(  S\right)  }\gamma_{k}\left(  \psi^{-1}\left(  k\right)
\cap S\right)  $, achieved by the undivided $S$. We conclude that $S$ can
guarantee \emph{at least} the payoff $\sum_{k\in\psi\left(  S\right)  }%
\gamma_{k}\left(  \psi^{-1}\left(  k\right)  \cap S\right)  $ to itself. But
\emph{one} of the courses of action available to $N\diagdown S$ is to walk
away from $S$ and access facilities on its own, which implies that $S$ can
guarantee \emph{at most} $\sum_{k\in\psi\left(  S\right)  }\gamma_{k}\left(
\psi^{-1}\left(  k\right)  \cap S\right)  .$ The upshot is that $v\left(
S\right)  =\sum_{k\in\psi\left(  S\right)  }\gamma_{k}\left(  \psi^{-1}\left(
k\right)  \cap S\right)  $, as in our Definition \ref{formula for v}.

Note that, in the model above, a coalition $S$ is deemed to have the strategic
ability to partition itself into $S_{1},\ldots,S_{l}$ once for all,
\emph{unconditional on} $k\in K$. When $S_{i}$ is reported to $\left(
\psi,\gamma\right)  ,$ the prerogative belongs to $\psi$ to determine the
user-sets $\psi^{-1}\left(  k\right)  \cap S_{i}$ at $k\in K$. The coalition
$S$ \emph{cannot} strategically create different partitions of itself
conditional on $k\in K$, indeed we may even suppose it does not know the set
$K$ of facilities\footnote{Otherwise, even in our canonical case of
$\mathcal{C=N}$ and user-blind facilities with constant costs $\gamma_{k}$ for
all $k\in K$, any coalition $S$ would split itself into singletons at $k$ for
which $\gamma_{k}>0$ and go undivided as $S$ at $k$ for which $\gamma_{k}<0$
(and the formula (\ref{Game Rep 1}) in Section \ref{representations}\ would be
invalid). This is ruled out in our model. Any coalition $S\in\mathcal{C}$ may
approach $\left(  \psi,\gamma\right)  $ as a \textquotedblleft black
box\textquotedblright\ without insider information of $\psi$ or $\gamma$, in
particular not even information of the set $K$ (which is the range of $\psi$
and the domain of $\gamma).$ What $S$ can query the black box about is the
payoff that will be forthcoming to any coalition $T\in\mathcal{C}$; and the
black box responds with the number $\sum_{k\in\psi(T)}\gamma\left(  k\right)
$. Thus $S$ can only acquire the incomplete information about $\left(
\psi,\gamma\right)  $ that is summarized in the game $v_{\left(  \psi
,\gamma\right)  }$ induced by $\left(  \psi,\gamma\right)  .$ Based upon this
information, it must decide on the partition $S=S_{1}\cup\ldots\cup S_{l}$ to
split itself into.}.

Further note that the $MM$ games $w_{S}$ (see Definition \ref{MM games}) are
\emph{not }superadditive, but $-w_{S}$ are, and we could have used $-w_{S}$
instead of $w_{S}$ throughout our analysis, since Span$\left\{  w_{S}%
:S\in\mathcal{C}\right\}  =$Span$\left\{  -w_{S}:S\in\mathcal{C}\right\}  .$

The set of superadditive games%
\[
\mathcal{V}\left(  \mathcal{C}\right)  =\left\{  v\in\mathbb{R}^{\mathcal{C}%
}:v\text{ is superadditive}\right\}
\]
is a full-dimensional cone\footnote{A set of games is called a
\textquotedblleft cone\textquotedblright\ if it is closed under non-negative
linear combinations.} in the Euclidean space $\mathbb{R}^{\mathcal{C}}$ of all
games, and Theorem \ref{necessity of full span} is easily checked to hold with
$\mathcal{V}\left(  \mathcal{C}\right)  $ in place of $\mathbb{R}%
^{\mathcal{C}}.$ (Note that it is only in Theorem \ref{necessity of full span}
that we refer to a \emph{space} of games, while the rest of our analysis holds
for any game $v\in\mathbb{R}^{\mathcal{C}}$ whatsoever, whether superadditive
or not.)

In fact, Theorem \ref{necessity of full span} holds also for the
full-dimensional sub-cone $\mathcal{V}^{\ast}\left(  \mathcal{C}\right)  $ of
$\mathcal{V}\left(  \mathcal{C}\right)  ,$ consisting of superadditive
\emph{cost-allocation games}, given by\emph{ }%
\[
\mathcal{V}^{\ast}\left(  \mathcal{C}\right)  =\left\{  v\in\mathcal{V}\left(
\mathcal{C}\right)  :v\leq0\right\}  .
\]
When $v\in\mathcal{V}^{\ast}\left(  \mathcal{C}\right)  $ we \emph{can} allow
any coalition $S\in\mathcal{C}$ to partition itself differentially,
conditional on $k\in K$ in our model. However $S$ will choose, in its own
interest, to remain undivided at each $k$. Thus complete information of
$\left(  \psi,\gamma\right)  ,$ or incomplete information of $\left(
\psi,\gamma\right)  $ summarized by its induced game $v_{\left(  \psi
,\gamma\right)  },$ become equivalent when $v\in\mathcal{V}^{\ast}\left(
\mathcal{C}\right)  .$

\subsection{A Class of Discriminatory Facilities}

We now present a heuristic argument as to why discriminatory facilities are
not as conducive to our analysis as user-blind facilities. For convenience, we
shall restrict attention to a particular class of discriminatory facilities.

Consider an \textquotedblleft$(S,m,c)$\emph{-facility}\textquotedblright%
\ $k=k\left(  S,m,c\right)  $, based\ on an exogenously specified coalition
$S\in\mathcal{C}$, a positive integer $1\leq m\leq\left\vert S\right\vert $
and a scalar $c\neq0$, which operates as follows. The cost $\gamma_{k}\left(
T\right)  $ to any set $T\subset N$ of users of $k$ is%

\begin{equation}
\gamma_{k}\left(  T\right)  =\left\{
\begin{array}
[c]{ccc}%
c & \text{if} & \left\vert S\cap T\right\vert \geq m\\
0 & \text{if} & \left\vert S\cap T\right\vert <m
\end{array}
\right.  \label{observant facility}%
\end{equation}
(Note that our user-blind facilities correspond to the case $m=0,$ and are not
in the picture here. Also, $\gamma_{k}\left(  \emptyset\right)  $ is
understood to be $0$ as usual.)

To begin with, let us focus on the \textquotedblleft$\left(  S,c\right)
$\emph{-facility}\textquotedblright\ $k=k\left(  S,c\right)  $ obtained by
setting $m=\left\vert S\right\vert $, where the condition \textquotedblleft%
$\left\vert S\cap T\right\vert \geq m$\textquotedblright\ reduces to
\textquotedblleft$S\subseteq T$\textquotedblright. In the spirit of our
earlier Definition \ref{defn symmetry} of symmetry (but now from the
\textquotedblleft local\textquotedblright\ $k$-standpoint in place of the
earlier \textquotedblleft global\textquotedblright\ standpoint of the abstract
game $v),$ two players $i$ and $j$ should be considered \emph{symmetric}
\emph{at} $k$ if their role is interchangeable in the map $\gamma
_{k}:\mathcal{K}_{k}\longrightarrow\mathbb{R}$ (see (\ref{computer program})).
Taking both the domain and the range of $\gamma_{k}$ into account, this means
that replacing $i$ with $j$ should makes for no difference in terms of the
\emph{membership in }$\mathcal{K}_{k},$ as well as the cost incurred at $k,$
i.e., for any $T\subseteq N$ with $i\notin T$ and $j\notin T$ (where $T$ could
be the empty set) we must have%
\begin{equation}
T\cup\left\{  i\right\}  \in\mathcal{K}_{k}\Longleftrightarrow T\cup\left\{
j\right\}  \in\mathcal{K}_{k} \label{subsitutes}%
\end{equation}
and%
\begin{equation}
\gamma_{k}\left(  T\cup\left\{  i\right\}  \right)  =\gamma_{k}\left(
T\cup\left\{  j\right\}  \right)  \text{ if }T\cup\left\{  i\right\}
\in\mathcal{K}_{k} \label{sub 2}%
\end{equation}
Note that if $\mathcal{C=N}$ and $\psi\left(  n\right)  =K$ for all $n\in N,$
then (\ref{subsitutes}) holds automatically and, in view of
(\ref{observant facility}), the requirement (\ref{sub 2}) reduces to: either
$i,j\in S$ or $i,j\notin S$ (in other words, two players are symmetric if, and
only if, both belong to $S$ or both are outside of $S).$

In the same vein (i.e., in the spirit of our earlier Definition
\ref{defn dummy}) a player $i$ should be considered a \emph{dummy }at $k$ if
$i$ is irrelevant both in terms of membership in\emph{ }$\mathcal{K}_{k}$ and
the cost incurred at $k$, i.e., for any $T\subseteq N$ with $i\notin T,$
\begin{equation}
T\cup\left\{  i\right\}  \in\mathcal{K}_{k}\mathfrak{\Longleftrightarrow}%
T\in\mathcal{K}_{k} \label{dummy}%
\end{equation}
and
\begin{equation}
\gamma_{k}\left(  T\cup\left\{  i\right\}  \right)  =\gamma_{k}\left(
T\right)  \text{ if }T\in\mathcal{K}_{k} \label{dummy 2}%
\end{equation}
If $\mathcal{C=N}$ then $i$ is a dummy if, and only if, $i\notin S$.

With these definitions, it does stand to reason (even in the non-canonical
case $\mathcal{C\subsetneq N}$) that symmetric players at facility $k$ should
bear equal portions of its cost, while dummies at $k$ should bear none of it.

Now consider a representation $\left(  \psi,\gamma\right)  $ of $v\in
\mathbb{R}^{\mathcal{C}}$ all of whose facilities are $\left(  S,c_{S}\right)
$-facilities for\footnote{Since we must have $\psi^{-1}\left(  k\right)
\in\mathcal{C}$ in any representation $\left(  \psi,\gamma\right)  $ of
$v\in\mathcal{R}^{C},$ it follows that we must restrict to $S$-facilities
where $S\in\mathcal{C}$.} $S\in\mathcal{C}$ (we write $c_{S}$ to allow costs
to vary with $S$). The key question is: \emph{under what conditions on
}$\mathcal{C}$ (\emph{and an appropriately constructed }$\psi$) \emph{can the
players in }$N$ \emph{be partitioned into two sets }$A_{S}$\emph{ and }$B_{S}%
$\emph{, where }$A_{S}$ \emph{consists of symmetric players and }$B_{S}$
\emph{of dummies, at every }$\left(  S,c_{S}\right)  $\emph{-facility? For
only in this case would it be justified to split the cost }$c_{S}$\emph{
equally among the players in }$A_{S}$ (after which one could aggregate the
cost-allocation across the decentralized facilities to arrive at the overall allocation).

In the canonical case $\mathcal{C}=\mathcal{N}$, the answer is: always (which
enables us to use such discriminatory facilities in our game representations,
just as well as user-blind facilities.) Indeed, for each $S\in\mathcal{N}$,
recall that the \textquotedblleft unanimity game\textquotedblright%
\ $u_{S}=\left(  u_{S}\left(  T\right)  \right)  _{T\in\mathcal{N}}%
\in\mathbb{R}^{\mathcal{N}}$ is given by%
\[
u_{S}\left(  T\right)  =\left\{
\begin{array}
[c]{ccc}%
1 & \text{if} & S\subseteq T\\
0 & \text{if} & S\nsubseteq T
\end{array}
\right.
\]
As is well-known since \cite{Shapley 1951} and \cite{Shapley 1953}, $\left\{
u_{S}\right\}  _{S\in\mathcal{N}}$ constitutes a basis of $\mathbb{R}%
^{\mathcal{N}}$. Hence, for any $v\in\mathbb{R}^{\mathcal{N}}$, there exist
unique scalars $d_{S}$ such that
\[
v=%
{\displaystyle\sum\nolimits_{S\in\mathcal{N}}}
d_{S}u_{S}%
\]
For each $S\in\mathcal{N}$ consider the discriminatory $\left(  S,d_{S}%
\right)  $-facility. Denote (without confusion) the set of $S$-facilities also
by $\mathcal{N},$ and put $\widetilde{\psi}\left(  n\right)  =\mathcal{N}$ for
$n\in N$, i.e., each player must use all the facilities. Also denote by
$\widetilde{\gamma}_{k}$ the family of maps generated by the $d_{S}$, i.e.,
given by (\ref{observant facility}) with $c=d_{S}$ when $k$ corresponds to
$S.$ (Notice that the map $\widetilde{\psi}$ is non-discriminatory in $\left(
\widetilde{\psi},\widetilde{\gamma}\right)  $ while the facilities are
discriminatory, which is just the opposite of what happened in our
representation with user-blind facilities.)

One may readily verify that:

(i) $\left(  \widetilde{\psi},\widetilde{\gamma}\right)  $ is a representation
of $v;$

(ii) players in $S$ are symmetric, and players in $N\diagdown S$ are not just
symmetric but dummies, at each $\left(  S,d_{S}\right)  $-facility (conditions
(\ref{subsitutes}),(\ref{sub 2}) and (\ref{dummy}),(\ref{dummy 2}) hold
automatically when $\mathcal{C}=\mathcal{N}$);

(iii) the local game induced by $\left(  \widetilde{\psi},\widetilde{\gamma
}\right)  $ at each $\left(  S,d_{S}\right)  $-facility is the game
$d_{S}u_{S};$

(iv) the equitable solution of $\left(  \widetilde{\psi},\widetilde{\gamma
}\right)  ,$ defined by splitting the cost $d_{S}$ equally among the
substitute players in $S$ at each $\left(  S,d_{S}\right)  $-facility for
$S\in\mathcal{N}$, yields the Shapley value of the game $v.$

\emph{The above construction, with discriminatory }$\left(  S,d_{S}\right)
$-\emph{facilities in the canonical case }$\mathcal{C}=\mathcal{N},$
\emph{does not extend when we step outside the canon into the new territory
}$\mathcal{C}\subsetneqq\mathcal{N}$ that we are exploring. The difficulty
arises as follows. For $S\in\mathcal{C}$, define the \emph{unanimity game
}$v_{S}\in\mathbb{R}^{\mathcal{C}}$ to be the restriction to $\mathcal{C}$ of
the unanimity game\ $u_{S}\in\mathbb{R}^{\mathcal{N}}$ on $\mathcal{N}$, i.e.,%
\begin{equation}
v_{S}\left(  T\right)  =u_{S}\left(  T\right)  \text{ for }T\in\mathcal{C}
\label{unanimity game}%
\end{equation}

Suppose that $v^{\prime}\in\mathbb{R}^{\mathcal{C}}$ is in the span of
$\left\{  v_{S}:S\in\mathcal{C}\right\}  $ so that we may write $v^{\prime}=%
{\displaystyle\sum\nolimits_{S\in\mathcal{C}}}
\alpha_{S}v_{S}$ for suitable scalars $\alpha_{S}.$ Any representation
$\left(  \psi^{\prime},\gamma^{\prime}\right)  $ of $v^{\prime}$, with
discriminatory $\left(  S,\alpha_{S}\right)  $-facilities\footnote{$\bigskip
$i.e., the cost function at the $\left(  S,\alpha_{S}\right)  $-facility is
given by:
\par
$\gamma_{S}\left(  T\right)  =\left\{
\begin{array}
[c]{ccc}%
\alpha_{S} & \text{if} & S\subset T\\
0 & \text{otherwise} &
\end{array}
\right.  $} for $S\in\mathcal{C}$, must\footnote{Recall that the
representation $\left(  \psi^{\prime},\gamma^{\prime}\right)  $ of $v^{\prime
}$ must fulfil conditions (\ref{Game Rep 1}) and (\ref{Game Rep 2}). Condition
(\ref{Game Rep 1}) is automatic because $v^{\prime}=%
{\displaystyle\sum\nolimits_{S\in\mathcal{C}}}
c_{S}v_{S}.$ Condition (\ref{Game Rep 2}) is not automatic, but let us assume
that that it also holds (e.g., assume $N\in\mathcal{C}$ and $\psi\left(
n\right)  =\mathcal{C}$ for all $n\in N$, analogous to the canonical case).
The problem that we are pointing out is something else, it goes well beyond
(\ref{Game Rep 2}).} induce the local game $\alpha_{S}v_{S}$ at each $\left(
S,\alpha_{S}\right)  $-facility; and this is easily achieved by again setting
$\psi^{\prime}\left(  n\right)  =K=\mathcal{C}$ for every player $n\in N.$
However \emph{it is no longer the case, without further ado, that the players
can be partitioned into symmetric players }$A_{S}$ \emph{and dummies }$B_{S}$
\emph{at these facilities, in accordance with (\ref{subsitutes}),(\ref{sub 2})
and (\ref{dummy}),(\ref{dummy 2}). }(It is (\ref{subsitutes}) and
(\ref{dummy}) that fail to hold.) Thus we are left in the dark as to how to
allocate the cost $\alpha_{S}$ among the players in $N$ at the $\left(
S,\alpha_{S}\right)  $-facility.

For such a partition to exist, many more conditions on $\mathcal{C}$ will be
needed in general, over and above the fact that $v^{\prime}$ is in the span of
$\left\{  v_{S}:S\in\mathcal{C}\right\}  $. One could of course
\emph{postulate }the equal split of the cost $\alpha_{S}$ among $S$ at every
discriminatory $\left(  S,\alpha_{S}\right)  $-facility\footnote{equivalently,
in each local game $c_{S}v_{S}$ induced by $\left(  \widetilde{\psi
},\widetilde{\gamma}\right)  $ at the $S$-facility}, and then extend these
cost allocations additively to an overall allocation in $\left(  \psi^{\prime
},\gamma^{\prime}\right)  $, i.e., in the game $v^{\prime}$. But such a
postulate would be totally arbitrary in the absence of the $A_{S},B_{S}$
partitions. On the other hand, for such partitions to exist, stringent
conditions must be placed $\mathcal{C}$ in addition to full span, and it is
not clear (at least to the author) if these can be stated in any simple form.
(One \emph{very} stringent condition on $\mathcal{C}$ that would suffice is as
follows: for any coalition $S\in\mathcal{C}$ and $i\in S$, the membership of
$S\diagdown\left\{  i\right\}  $ in $\mathcal{C}$ is invariant of the choice
of $i.$) Thus discriminatory facilities do not take us very far in our analysis.

\subsubsection{Example 3 (User-Blind vs Discriminatory Facilities)}

It might help to consider an example that reveals the difference between
user-blind versus discriminatory facilities. Consider the cost-allocation game
$v$ on the semi-algebra of coalitions $\left\{  \left\{  1\right\}  ,\left\{
23\right\}  ,\left\{  123\right\}  \right\}  $ given by $v\left(  1\right)
=-1,$ $v(23)=-2,$ $v\left(  123\right)  =-4.$ With user-blind facilities, it
is readily verified that the unique minimal representation of $v$ involves
three facilities $k_{1},k_{2},k_{3}$ (corresponding to $\left\{  1\right\}
,\left\{  23\right\}  ,\left\{  123\right\}  $) and is given by $\left(
\psi,\gamma\right)  ,$ where $\psi\left(  1\right)  =\left\{  k_{1}%
,k_{3}\right\}  ,\psi\left(  2\right)  =\psi\left(  3\right)  =\left\{
k_{2},k_{3}\right\}  $ and $\gamma\left(  k_{1}\right)  =-2,$ $\gamma\left(
k_{2}\right)  =-3,$ $\gamma\left(  k_{3}\right)  =1$. Thus the equitable
solution, based on splitting costs equally among users at each facility, is
$\left(  -5/3,-7/3,-7/3\right)  .$ Incidentally, the local games induced at
$k_{1},k_{2},k_{3}$ are $\left(  -2,0,-2\right)  $, $\left(  0,-3,-3\right)
,$ $\left(  1,1,1\right)  $ with the first,second,third component
corresponding to the payoffs of coalitions $\left\{  1\right\}  ,\left\{
23\right\}  ,\left\{  123\right\}  $ respectively. The sum of the three local
games gives $v=\left(  -1,-2,-4\right)  $ which verifies that we have a bona
fide representation. (See (\ref{sum of local games}).)

The representation $\left(  \psi^{\prime},\gamma^{\prime}\right)  $ based on
discriminatory $\left(  S,c_{S}\right)  $-facilities is as follows:
$\psi^{\prime}\left(  n\right)  =\left\{  k_{1},k_{2},k_{3}\right\}  $ for
$n=1,2,3$; and the cost functions $\gamma_{k_{i}}$ are given by
(\ref{observant facility}) with $\left(  S,c_{S}\right)  $ set equal to
$\left(  \left\{  1\right\}  ,-1\right)  $ at $k_{1},$ $\left(  \left\{
2,3\right\}  ,-2\right)  $ at $k_{2}$ and $\left(  \left\{  1,2,3\right\}
,-1\right)  $ at $k_{3}$ $.$ The local games induced at $k_{1},k_{2},k_{3}$
are now $\left(  -1,0,-1\right)  $, $\left(  0,-2,-2\right)  ,$ $\left(
0,0,-1\right)  $ whose sum is also $v=\left(  -1,-2,-4\right)  $ which
verifies (again by (\ref{sum of local games})) that the representation is bona
fide. However the equal-split solution now yields the cost allocation $\left(
-4/3,-4/3,-4/3\right)  $ which quite different from $\left(
-5/3,-7/3,-7/3\right)  .$

Note that $K_{k_{3}}=\left\{  \left\{  1\right\}  ,\left\{  23\right\}
,\left\{  123\right\}  \right\}  $ in $\left(  \psi^{\prime},\gamma^{\prime
}\right)  ,$ thus condition (\ref{subsitutes}) does not hold for $1$,$2$ which
implies that $1$ and $2$ are not symmetric and nor are $1$ and $3,$ though $2$
and $3$ are symmetric. Thus the equal split of the cost at $k_{3}$ among
players $1,2,3$ is not justified, which accounts for the difference between
the two solutions.

\begin{remark}
\textbf{(Dual Games) }Suppose $\mathcal{C}$ is a semi-algebra. Then the
\emph{dual }$v^{\ast}\in\mathbb{R}^{\mathcal{C}}$ of a game $v\in
\mathbb{R}^{\mathcal{C}}$ can be defined by $v^{\ast}\left(  S\right)
=v\left(  N\right)  -v\left(  N\diagdown S\right)  $\footnote{As usual
$v\left(  \emptyset\right)  $ is understood to be $0$ here. Also observe that
$v^{\ast\ast}=v$, i.e., the dual of the dual of $v$ is $v.$ (Dual games have
been examined in the canonical case $\mathcal{C=N}$, see e.g. \cite{Bilbao}
and the references therein.)}. It is readily verified that $MM$-games and
unanimity games are duals of each other, i.e., $w_{S}^{\ast}=v_{S}$ and
$v_{S}^{\ast}=w_{S}$ for any $S\in\mathcal{C}$; and that $\left\{
w_{S}\right\}  _{S\in\mathcal{C}}$ and $\left\{  v_{S}\right\}  _{S\in
\mathcal{C}}$ each constitute a basis of $\mathbb{R}^{\mathcal{C}}.$ Thus
representations of $v$ can be constructed with user-blind (resp.,
discriminatory) $\left(  S,\alpha_{S}\right)  $-facilities, where the local
games induced at the facilities are -- upto scalar multiplication --
$MM$-games (resp., unanimity games). Equal-split of costs at these facilities
will lead to an allocation in $v$ in either case, but the two allocations will
typically be different, as our example illustrates. The burden of our song (we
reiterate) is that equal split is not justified for discriminatory facilities
since conditions (\ref{subsitutes}) and (\ref{dummy}) fail to hold without
additional stringent conditions on $\mathcal{C};$ whereas it is justified for
user-blind facilities since those conditions become irrelevant. (Recall that
the conditions pertain to scrutiny of user-sets that come to a facility, and
such scrutiny can be regarded as missing in the case of user-blind facilities.)
\end{remark}

\subsection{Conclusion}

The foregoing discussion extends \textit{mutatis mutandis} to the case of
general $1\leq m<\left\vert S\right\vert .$ In the canonical case
$\mathcal{C}=\mathcal{N}$, one can (by varying $m=m\left(  S\right)  $ across
$S\in\mathcal{N}$) generate \emph{all} bases $\left\{  w_{S}:S\in
\mathcal{N}\right\}  $ of $\mathbb{R}^{\mathcal{N}}$ that fulfil the three
\textquotedblleft standard\textquotedblright\ provisos: (i) each $w_{S}$ is
monotonic; (ii) all players in $S$ are symmetric in $w_{S}$; (iii) all players
in $N\diagdown S$ are dummies in $w_{S}$. And it does not matter which basis
we use as a foundation for the game representation, because any such
representation will induce back the local game $w_{S}$ at the facility labeled
$S\in\mathcal{N}$, and thus the lead to the Shapley value. However when
$\mathcal{C}\subsetneqq\mathcal{N}$\emph{ }we are beset with the same problem
that arose in the special case $m=\left\vert S\right\vert $, since it is no
longer true that the players in $S$ are symmetric and those outside $S$ are
dummies, in accordance with (\ref{subsitutes}),(\ref{sub 2}) and
(\ref{dummy}),(\ref{dummy 2}), at the discriminatory $\left(  S,m,c\right)
$-facilities without stringent conditions on $\mathcal{C}$ over and above full
span condition.\footnote{The full span condition here is, of course, meant to
hold in terms of the local games induced at the $\left(  S,m,c\right)
$-facilities, not in terms of the $MM$-games defined earlier that were induced
by user-blind facilities.}

Thus it is at the heart of our analysis that facilities be \emph{user-blind}%
\footnote{They correspond to $(S,m,c)$-facilities with $m=0,$ which is
tantamount to saying that every user-set is charged the same, i.e., the
facility is user-blind and conditions \emph{(}\ref{subsitutes}),(\ref{sub 2})
and (\ref{dummy}),(\ref{dummy 2}) are rendered irrelevant.} in our
representations$.$

Finally, if it \emph{does} turn out that players can\emph{ }be partitioned
into symmetric players and dummies at some discriminatory $\left(
S,m,c\right)  $-facility, then our equitable solution $\chi\left(  w\right)  $
(see Section \ref{Ext Shapley Value}) of the game $w$ induced at this $\left(
S,m,c\right)  $-facility will also divide the entire cost of the facility
equally among the symmetric players and charge nothing to the dummies. In
short, the introduction of discriminatory facilities --- on the limited
occasions where they can be useful --- yields nothing new. This is implied by
the following Claim, which follows immediately from Theorem \ref{properties}.

\begin{claim}
\label{claim}Suppose $\mathcal{C}$ satisfies full span and $\left(
\psi,\gamma\right)  $ is a representation of $v\in\mathbb{R}^{\mathcal{C}}$
which has some discriminatory $\left(  S,m,c\right)  $-facility. Suppose
further that the players in $N$ can be partitioned into symmetric and dummies
at the $\left(  S,m,c\right)  $-facility in accordance with
\emph{(\ref{subsitutes}),(\ref{sub 2}) }and \emph{(\ref{dummy}),(\ref{dummy 2}%
)}. Then the equitable solution $\chi\left(  w\right)  $ of the game $w$
induced at the $\left(  S,m,c\right)  $-facility divides the total cost of the
facility equally among all the symmetric and charges nothing to the dummies.
\end{claim}

\section{Comments\label{comments}}

(1) \textbf{(Minimal Facilities)} It is evident from the proof of Theorem
\ref{general TU} that, for generic\footnote{Precisely: for all $v\in
\mathcal{V}\left(  N\right)  ,$ after deleting a fiinite number of
lower-dimensional subspaces.} $v\in\mathbb{R}^{C}$ , the unique minimal
representation of $v$ needs $\left\vert \mathcal{C}\right\vert $ facilities.

(2) (\textbf{Necessity of Condition (}\ref{Game Rep 2})\textbf{) }Consider any
$w$ on the semi-algebra $\left\{  \emptyset,S,N\diagdown S,N\right\}  $ and
extend it arbitrarily to $v$ on $\mathcal{N}.$ Any representation of $v$ will
also serve as a representation of $w$ if condition (\ref{Game Rep 2}) were to
be dropped, but the equitable solution of $v$ varies widely depending on the
extension $v$ that is chosen. This demonstrates the necessity of condition
(\ref{Game Rep 2}) in the definition of representations, given in Section
\ref{representations}.

(4) (\textbf{A Formula for the Extension }$\chi$ \textbf{of the Shapley
Value)} Assume the full span condition on $\mathcal{C}$. Then any
$v\in\mathbb{R}^{\mathcal{C}}$ has a unique linear expansion in terms of the
MM-basis $\left\{  w_{S}\right\}  _{S\in\mathcal{C}}$. The coefficients in the
expansion, which give the costs $\gamma$ of the facilities in the minimal
representation of $v,$ are obtained by inverting the matrix $A$ associated
with $\left\{  w_{S}\right\}  _{S\in\mathcal{C}}$ (see Remark \ref{matrix}),
indeed $\gamma=A^{-1}v.$ Then it is a simple matter to carry out equal
division of costs at each facility to get the equitable solution $\chi\left(
v\right)  .$ This gives a formula for $\chi\left(  v\right)  $ in terms of $v$
and $A.$

(5) (\textbf{An Equivalent Game\footnote{This comment is due to Ori Haimanko.}
on }$\mathcal{N}$\textbf{) }Assume the full span condition on $\mathcal{C}$
and consider any $v\in\mathbb{R}^{C}.$ There exists a full-blown game
$\widetilde{v}\in\mathbb{R}^{\mathcal{N}}$ such that $\chi\left(  v\right)
=\varphi\left(  \widetilde{v}\right)  ,$ i.e., the equitable solution of $v$
is the Shapley value of $\widetilde{v}.$ To see this, write $v=%
{\textstyle\sum_{S\in\mathcal{C}}}
\alpha_{S}w_{S}$ where $\left\{  w_{S}\right\}  _{S\in\mathcal{C}}$ is the
MM-basis of $\mathbb{R}^{C}$ and the $\alpha_{S}$ are (uniquely determined)
scalars. Extend each vector $w_{S}\in\mathbb{R}^{\mathcal{C}}$ to a vector
$\widetilde{w}_{S}\in\mathbb{R}^{\mathcal{N}}$ by setting
\[
\widetilde{w}_{S}\left(  T\right)  =\left\{
\begin{array}
[c]{ccc}%
1 & \text{if} & S\cap T\neq\emptyset\\
0 & \text{if} & S\cap T=\emptyset
\end{array}
\right.
\]
for all $T\in\mathcal{N}$. Define the game $\widetilde{v}\in\mathbb{R}%
^{\mathcal{N}}$ by $\widetilde{v}=%
{\textstyle\sum_{S\in\mathcal{C}}}
\alpha_{S}\widetilde{w}_{S}.$ We claim that $\chi\left(  v\right)  =%
{\textstyle\sum_{S\in\mathcal{C}}}
\chi\left(  \alpha_{S}w_{S}\right)  =%
{\textstyle\sum_{S\in\mathcal{C}}}
\varphi\left(  \alpha_{S}\widetilde{w}_{S}\right)  =\varphi\left(
\widetilde{v}\right)  $. The first equality follows from the additivity
property of $\chi,$established in Theorem \ref{properties}. The last equality
follows from the (well-known) additivity property of the Shapley value.
Finally note that, for every $S\in\mathcal{C}$, the players in $S$ are
symmetric while those in $N\diagdown S$ are dummies in the game $\alpha
_{S}\widetilde{w}_{S}$ and that $\left(  \alpha_{S}\widetilde{w}_{S}\right)
(N)=\alpha_{S}.$ Therefore by the (also well-known) efficiency, dummy, and
substitute properties of the Shapley value, we conclude that $\varphi\left(
\alpha_{S}\widetilde{w}_{S}\right)  $ divides $\alpha_{S}$ equally among the
players in $S$ and gives $0$ to the players in $N\diagdown S.$ But this is
exactly what the equitable solution $\chi$ awards the players in the game
$\alpha_{S}w_{S}\in\mathbb{R}^{C}.$ Thus $\chi\left(  \alpha_{S}w_{S}\right)
=\varphi\left(  \alpha_{S}\widetilde{w}_{S}\right)  $ for every $S\in
\mathcal{C}$, establishing the middle inequality.

However, this observation is not of much use in the computation of
$\chi\left(  v\right)  $ as it requires us to first compute the coefficients
$\alpha_{S}$ in the linear expansion $v=%
{\textstyle\sum_{S\in\mathcal{C}}}
\alpha_{S}w_{S}$ to get to $\widetilde{v};$ but we get $\chi\left(  v\right)
$ immediately from the $\alpha_{S}$ as pointed out in the previous remark,
without the additional labor of computing $\varphi\left(  \widetilde{v}%
\right)  .$

(6) \textbf{(Axiomatization of }$\chi$?) In view of Theorem \ref{properties},
it is tempting to think of an axiomatization of $\chi:\mathbb{R}%
^{C}\longrightarrow\mathbb{R}^{N}$ by means of Symmetry, Dummy and Linearity;
but this endeavor will not work for general $\mathcal{C}$ (even after assuming
the full span condition). The trouble is as follows. Consider the unique
minimal representation of a game $v\in\mathbb{R}^{C}$ whose user-blind
facilities are also indexed by $S\in C$ (for generic $v$, see Comment 1
above). The local game (see Definition \ref{defn local game}) induced at the
$S$-facility is $c_{S}w_{S}$ where $c_{S}$ is the cost of that facility and
$w_{S}$ is the $MM$-game of Definition \ref{MM games}. Now it is \emph{not}
generally the case that the players outside $S$ are dummies while those in $S$
are symmetric to each other in $c_{S}w_{S}$ in accordance with Definitions
\ref{defn dummy}, \ref{defn symmetry} (without imposing additional stringent
conditions on $\mathcal{C)}$. Therefore the equal-split of $c_{S}$ (among the
players in $S$) at the $S$-facility cannot be deduced from the Symmetry and
Dummy axioms applied to the game $c_{S}w_{S}.$ These axioms are simply
ineffectual in that game. It is the $k$-standpoint, in conjunction with the
user-blindness of $k$, that enables us to postulate the equal-split. (When
facilities are discriminatory, this rationale for the equal-split also breaks
down, since conditions (\ref{subsitutes}) and (\ref{dummy}) generally fail to
hold hold as was pointed out earlier.)

(7) (\textbf{Still} \textbf{Other Representations?) }Given a characteristic
function $v$ on $\mathcal{N}$, is it possible to conjure \textquotedblleft
representations\textquotedblright\ of $v$ in which \textquotedblleft intuitive
solution\textquotedblright\ are not equal to the Shapley value of $v$? In this
paper,\emph{ }a \textquotedblleft representation\textquotedblright\ is
\emph{defined} to be an assignment, and then the answer is no. The question
will become operational if a broader notion of \textquotedblleft
representations\textquotedblright, and of their \textquotedblleft intuitive
solutions\textquotedblright, can be made mathematical. The author has no clue
as to how to go about this.

\section{Related Literature}

By way of a concrete instance of the assignment $\left(  \psi,\gamma\right)
$, think of $N$ as a set of identical planes, and of $\psi(n)$ as the set of
cities that plane $n$ must fly to.\footnote{The daily flight path of each
plane is a loop which traverses the cities in $\psi(n)$ (in any order), taking
off and landing at each city in $\psi(n)$ exactly once.} If $\gamma(k)<0,$ it
represents the cost of building the runway at city $k$ (a runway, once built,
accomodates any number of planes since they are of identical make); and if
$\gamma(k)>0,$ one may think of it as a \textquotedblleft
subsidy\textquotedblright\ given by city $k$ (collectively to the planes that
visit its remote location). These costs (benefits) may vary across cities on
account of different costs of labor, land, location, etc.

This picture is based on \cite{Littlechild}, which focused attention on a
single airport used by different-sized planes; and on \cite{Dubey 1982} which
observed that the model of \cite{Littlechild}\ can be recast with
identical-sized planes that have different flight paths among multiple
airports. However the analysis in \cite{Dubey 1982} (as in \cite{Littlechild})
was restricted to the case of costs $\gamma(k)<0$, which severely limited the
class of characteristic functions that could be generated. The fact was missed
in \cite{Dubey 1982} that, by letting $\gamma(k)$ take on arbitrary values in
$\mathbb{R}$, one can generate all possible characteristic functions.

There are other approaches to extending the Shapley Value for games on
curtailed coalitions,.based on Harsanyi's method of \textquotedblleft
dividends" (\cite{Harsanyi}, \cite{Albizuri}, \cite{Masuya}), or certain
proportionality rules (\cite{Wilson}), or symmetric linear spaces of games
(\cite{Housman}). They are complementary to the method of \textquotedblleft
game representations\textquotedblright\ described in this paper. It would be
interesting to explore the similarities and differences between the different
approaches, but this is a topic we defer to future work.

\section{Proofs\label{proofs}}

\subsubsection{Proof of Theorem \ref{general TU}}

\begin{proof}
Consider $v\in\mathbb{R}^{\mathcal{C}}.$ Since $\left\{  w_{S}\right\}
_{S\in\mathcal{C}}$ is a basis of $\mathbb{R}^{\mathcal{C}},$ there exist
unique scalars $c_{S}$ such that%
\begin{equation}
v=\sum_{S\in\mathcal{C}}c_{S}w_{S} \label{=1}%
\end{equation}

To construct an assignment for $v,$ first let the set of facilities $K$
correspond to (be indexed by) $\mathcal{C}$, so that we may write
$K=\mathcal{C}.$ (In the rest of this proof, the symbol $S$ will be used to
denote either a facility or a coalition, but this should cause no confusion,
as the meaning of $S$ will always be clear from the context.)

Define the assignment $\left(  \psi,\gamma\right)  $ as follows:
\begin{equation}
\psi(n)=\left\{  S\in\mathcal{C}:n\in S\right\}  ,\text{ for }n\in N;
\label{=2}%
\end{equation}

and%
\[
\mathcal{\gamma}\left(  S\right)  =c_{S}\text{ for }S\in\mathcal{C}%
\]

It is readily verified that $\left(  \psi,\gamma\right)  $ represents $v.$
Indeed observe that for the facility $S\in\mathcal{C},$ our construction
implies $\psi^{-1}(S)=S\in\mathcal{C}$. Next note that, by (\ref{=1}), we
have
\[
v(T)=\sum_{S\in\mathcal{C}}c_{S}w_{S}(T)
\]

for any coalition $T\in\mathcal{C}.$ But $w_{S}(T)=1$ if, and only if, $S$
meets $T$; and $w_{S}(T)=0$ otherwise. Therefore%
\[
v(T)=\sum_{S\in\mathcal{C}:S\text{ meets }T}c_{S}%
\]

Since, by (\ref{=2}), the users in $T$ are (collectively) linked by $\psi$ to
precisely those facilities $S$ such that $S$ meets $T$, we have $\psi
(T)=\left\{  S\in\mathcal{C}:S\text{ meets }T\right\}  ,$ hence%
\[
v(T)=\sum_{S\in\psi(T)}c_{S}\text{ for all }T\in\mathcal{C}%
\]

This shows that $\left(  \psi,\gamma\right)  $ represents $v.$

Denote $\mathcal{C}^{\ast}=\left\{  S\in\mathcal{C}:c_{S}\neq0\right\}  $ and
let $\left(  \psi^{\ast},\gamma^{\ast}\right)  $ be obtained from $\left(
\psi,\gamma\right)  $ by removing the facilities that have zero cost, i.e.,
restricting to facilities in $\mathcal{C}^{\ast}.$ We submit that (i) $\left(
\psi^{\ast},\gamma^{\ast}\right)  $ is the unique representation of $v$ with
the minimum number of facilities; (ii) all other representations are trivial
expansions of $\left(  \psi^{\ast},\gamma^{\ast}\right)  $. Indeed (i) follows
from (ii), so it suffices to prove (ii).

Let $\left(  \psi_{\#},\gamma_{\#}\right)  $ be an arbitrary representation of
$v$ with the set $K_{\#}$ of facilities. (Recall that the underlying set $N$
of users is held fixed throughout.) At each facility $k\in K_{\#}$ consider
the \textquotedblleft local game\textquotedblright\ $v^{k}\in\mathbb{R}%
^{\mathcal{C}}$ induced by $\left(  \psi_{\#},\gamma_{\#}\right)  ,$ i.e.,
\[
v^{k}\left(  T\right)  =\left\{
\begin{array}
[c]{ccc}%
\gamma_{\#}(k) & \text{if} & \psi_{\#}^{-1}(k)\text{ meets }T\\
0 & \text{if} & \psi_{\#}^{-1}(k)\text{ misses }T
\end{array}
\right.
\]

for all coalitions $T\in\mathcal{C}$. A moment's reflection reveals that%
\begin{equation}
v=\sum_{k\in K_{\#}}v^{k} \label{=3}%
\end{equation}

and that
\[
v^{k}=\gamma_{\#}(k)w_{S(k)}\text{ , where we denote }S(k)=\psi_{\#}%
^{-1}(k)\in\mathcal{C}%
\]

(The inclusion $\in$ holds above for the representation $\left(  \psi
_{\#},\gamma_{\#}\right)  ,$ by assumption). This implies
\[
v=\sum_{k\in K_{\#}}\gamma_{\#}(k)w_{S(k)}%
\]

From the fact that the linear expansion $v=\sum_{S\in\mathcal{C}}c_{S}w_{S}$
given in (\ref{=1}) is unique, we conclude that there is a partition $\left\{
K_{\#}(S):S\in\mathcal{C}\right\}  $ of $K_{\#}$ into equivalence classes of
the relation $\sim$ on $K_{\#}$ defined by: $k\sim l$ iff $S\left(  k\right)
=S\left(  l\right)  ,$ and that%
\[
v=\sum_{S\in\mathcal{C}}\left(  \sum_{k\in K_{\#}(S)}\gamma_{\#}(k)\right)
w_{S}%
\]

with%
\[
c_{S}=\sum_{k\in K_{\#}(S)}\gamma_{\#}(k)
\]

Thus the arbitrary representation $\left(  \psi_{\#},\gamma_{\#}\right)  $ of
$v$ is a trivial expansion of $\left(  \psi^{\ast},\gamma^{\ast}\right)  $.

Finally consider the canonical case $\mathcal{C=N}$ and let $\varphi
:\mathbb{R}^{\mathcal{N}}\longrightarrow\mathbb{R}^{N}$ denote the Shapley
value. Consider an arbirary representation $\left(  \psi_{\#},\gamma
_{\#}\right)  $ of $v$ with facilities $K_{\#}$. As in (\ref{=3}), let
$v=\sum_{k\in K_{\#}}v^{k}$ where the $v^{k}$ is the local game at facility
$k$ induced by $\left(  \psi_{\#},\gamma_{\#}\right)  $. Then $\varphi
(v)=\sum_{k\in K_{\#}}\varphi(v^{k})$ by the additivity of $\varphi.$ But, in
the game $v^{k},$ all the players in $\psi_{\#}^{-1}\left(  k\right)  $ are
symmetric and all the players in $N\backslash\psi_{\#}^{-1}\left(  k\right)  $
are dummies. Hence $\varphi(v^{k})$ divides the total $v^{k}(N)=\gamma
_{\#}(k)$ equally among the players in $\psi_{\#}^{-1}\left(  k\right)  $ and
gives $0$ to the players in $N\backslash\psi_{\#}^{-1}\left(  k\right)  $
(using the dummy, symmetry and efficiency properties of $\varphi).$ This
implies that $\varphi(v)$ is the equitable solution $\tau\left(  \psi
_{\#},\gamma_{\#}\right)  .$
\end{proof}

\subsubsection{Proof of Theorem \ref{Lemma}}

\begin{proof}
Let the sequence $S_{1},\ldots,S_{l}$ be a hierarchy for $\mathcal{C}$. For
any $S\in\mathcal{C},$ define $\eta_{S}\in\mathbb{R}^{\mathcal{C}}$ by%
\[
\eta_{S}(T)=1\text{ if }T=S\text{ , and }\eta_{S}(T)=0\text{ otherwise}%
\]

for all $T\in\mathcal{C}$, i.e., $\left\{  \eta_{S}:S\in\mathcal{C}\right\}  $
is the standard basis of $\mathbb{R}^{\mathcal{C}}$ consisting of the
\textquotedblleft unit\textquotedblright\ vectors. To show that $\mathcal{C}$
has full span, it suffices to show that each $\eta_{S}\in$ Span $\mathcal{W},$
where $\mathcal{W}=\left\{  w_{S}:S\in\mathcal{C}\right\}  .$

Consider $S_{l}$. Then there exists $T\in\mathcal{C}$ such that $T$ misses
$S_{l}$ and $T$ meets $S_{1},\ldots,S_{l-1}.$ But then it is evident that%
\[
\eta_{S_{l}}=w_{S_{1}}-w_{T}%
\]

i.e., $\eta_{S_{l}}\in$ Span $\mathcal{W}$.

Next assume inductively that $\eta_{S_{j}}\in$ Span $\mathcal{W}$ for all
$j\geq k$, for some $k\geq3.$ Consider $S_{k-1}$. Then there exists
$T\in\mathcal{C}$ such that $T$ misses $S_{k-1}$ and $T$ meets $S_{1}%
,\ldots,S_{k-2}.$ Thus the components $w_{S_{1}}\left(  U\right)  -w_{T}(U)$
of the vector $w_{S_{1}}-w_{T}$ have the following pattern: they are $0$ for
all $U=S_{1},\ldots,S_{k-2};$ and $1$ for $U=S_{k-1};$ and, finally, for each
$U=S_{k},\ldots,S_{l}$, they are obviously either $1$ or $0$ (depending on
whether $T$ misses $U$ or $T$ meets $U$). Denote%
\[
\mathcal{J=}\left\{  j:w_{S_{1}}\left(  U\right)  -w_{S_{j}}(U)=1,\text{and
}l\geq j\geq k\right\}
\]

Then it is clear that
\[
\eta_{S_{k-1}}=w_{S_{1}}-w_{T}-\sum_{j\in\mathcal{J}}\eta_{S_{j}}%
\]

Since $\eta_{S_{j}}\in$ Span $\mathcal{W}$ for every $j\in\mathcal{J}$ by the
inductive assumption, we conclude that $\eta_{S_{k-1}}\in$ Span $\mathcal{W}$.
This establishes that $\eta_{S_{j}}\in$ Span $\mathcal{W}$ for $j=2,\ldots,l.$
It only remains to check that $\eta_{S_{1}}\in$ Span $\mathcal{W}$. But that
follows from the observation%
\[
\eta_{S_{1}}=w_{S_{1}}-\sum_{j=2}^{l}\eta_{S_{j}}%
\]

We have proved the the second implication of the Lemma, i.e., if $\mathcal{C}$
is hierarchical, then $\mathcal{C}$ has full span.

Now suppose $\mathcal{C}$ is a semi-algebra. Let $0<k\left(  T\right)
<\ldots<k\left(  1\right)  =\left\vert N\right\vert $ be a list of the
cardinalities of \emph{all} the coalitions in $\mathcal{C}.$ Define%
\[
\mathcal{L}_{t}=\left\{  S\in\mathcal{C}:\left\vert S\right\vert =k\left(
t\right)  \right\}
\]

First note that $\mathcal{L}_{1}=\left\{  N\right\}  $ and that $N$ meets
every coalition in $\mathcal{C}$ since they are subsets of $N.$ Next, for any
$S\in\mathcal{L}_{t}$ where $t>1,$ consider its complement $S^{c}=N\diagdown
S.$ Then $S^{c}\in\mathcal{C}$ since $\mathcal{C}$ is a semi-algebra. Clearly
$S^{c}$ misses $S,$ and $S^{c}$ meets every coalition in $\mathcal{C}$ that is
not contained in $S.$ Denoting $\widetilde{\mathcal{L}}_{t}=\mathcal{L}%
_{1}\cup\ldots\cup\mathcal{L}_{t}$, note that all the coalitions in
$\widetilde{\mathcal{L}}_{t}\diagdown\left\{  S\right\}  $ have cardinality at
least that of $S$ and are distinct from $S,$ hence they are not contained in
$S.$ It follows that $S^{c}$ meets every coalition in $\widetilde{\mathcal{L}%
}_{t}\diagdown\left\{  S\right\}  .$

Now consider the sequence whose first element is $N,$ followed by all
coalitions of size $k\left(  2\right)  $ in an arbitrary order, then followed
by all coalitions of size $k\left(  3\right)  $ in an arbitrary order, and so
on ending with coalitions of size $k(T)$ in an arbitrary order. Our argument
in the previous paragraph shows that this sequence is a hierarchy for
$\mathcal{C}$, establishing the first implication of the lemma.
\end{proof}

\subsubsection{Proof of Theorem \ref{properties}}

\begin{proof}
Let
\[
v=\sum_{S\in\mathcal{C}}c_{S}w_{S}%
\]

be the unique expansion of $v$ in terms of the $MM$-basis. Suppose $i$ and $j$
are symmetric in $v$. Consider any $A\in\mathcal{C}$ with $i\in A$ and
$j\notin A.$ Then $B=(A\diagdown\left\{  i\right\}  )\cup\left\{  j\right\}
\in\mathcal{C}.$ We claim that $c_{A}=c_{B}.$ To see this, write%
\[
v=c_{A}w_{A}+c_{B}w_{B}+\sum c_{S}w_{S}%
\]

where the $\sum$ is over $S\in\mathcal{C\diagdown}\left\{  A,B\right\}  .$
Since $i$ and $j$ are symmetric, the above equation also holds if we swap
$c_{A}$ and $c_{B,}$ so%
\[
v=c_{B}w_{A}+c_{A}w_{B}+\sum c_{S}w_{S}%
\]

But the linear expansion is unique, so $c_{A}=c_{B}.$It follows that symmetric
players get the same payoff in the equitable solution of the unique minimal
representation\footnote{They clearly get the same at facilities indexed by
coalitions where both are absent or both are present; and, by virtue of what
we have just shown, also across the \emph{pair} of facilities where one is
present and the other absent.} of $v$;$.$and therefore, by Theorem
\ref{general TU}, this also holds in all representations of $v.$

Now dummies, being symmetric to each other, get the same payoffs also. To
verify that they in fact get $0$, take the minimal representation $\left(
\psi^{\ast},\gamma^{\ast}\right)  $ of $v$ and alter it as follows to obtain a
new assignment $\left(  \psi^{\prime},\gamma^{\prime}\right)  $. First remove
dummies from visiting every facility in the representation $\left(  \psi
^{\ast},\gamma^{\ast}\right)  $. Next add new facilities $k\left(  Q\right)  $
for every coalition $Q$ in $\mathcal{C}$ that contains dummies. Let the cost
of each $k\left(  Q\right)  $ be $0,$ and let visitors to facility $k\left(
Q\right)  $ be precisely the players in $Q.$ It is easily verified that (i)
$\left(  \psi^{\prime},\gamma^{\prime}\right)  $ is a representation of $v;$
(ii) the equitable solution gives $0$ to dummies in $\left(  \psi^{\prime
},\gamma^{\prime}\right)  $. Then, by Theorem \ref{general TU}, dummies get
$0$ in every representation of $v.$

To verify linearity, let $\left(  \psi^{v},\gamma^{v}\right)  $ and $\left(
\psi^{w},\gamma^{w}\right)  $ be representations of $v$ and $w$ with
facilities $K^{v}$ and $K^{w}$ respectively. Consider the assignment $\left(
\psi,\gamma\right)  $ obtained by concatenating $\left(  \psi^{v},\gamma
^{v}\right)  $ and $\left(  \psi^{w},\gamma^{w}\right)  ,$ i.e., the
facilities of $\left(  \psi,\gamma\right)  $ are the \emph{disjoint} union of
$K^{v}$ and $K^{w}$ (with costs in accordance with $\gamma^{v}$ and
$\gamma^{w}$)$,$ and $\psi\left(  n\right)  $ is the \emph{disjoint} union of
$\psi^{v}\left(  n\right)  $ and $\psi^{w}\left(  n\right)  $ for all $n\in
N.$ It is immediate that $\left(  \psi,\gamma\right)  $ is a representation of
$v+w$ and that $\tau\left(  \left(  \psi,\gamma\right)  \right)  =\tau\left(
\left(  \psi^{v},\gamma^{v}\right)  \right)  +\tau\left(  \left(  \psi
^{w},\gamma^{w}\right)  \right)  .$ By Theorem \ref{general TU}, it now
follows that $\chi\left(  v+w\right)  =\chi\left(  v\right)  +\chi\left(
w\right)  .$ Finally, if $\left(  \psi^{v},\gamma^{v}\right)  $ represents
$v,$ then clearly $\left(  \psi^{v},c\gamma^{v}\right)  $ represents $cv$,
which implies that $\chi\left(  cv\right)  =c\chi\left(  v\right)  .$
\end{proof}

\subsubsection{Proof of Theorem \ref{necessity of full span}}

\begin{proof}
It is readily verified that:

$\left(  \psi,\lambda\right)  $ is a representation of $v\in\mathbb{R}%
^{\mathcal{C}}$ with the set $K$ of facilities $\iff v=\sum_{k\in K}%
\lambda\left(  k\right)  w_{S\left(  k\right)  }$ where $S\left(  k\right)
=\psi^{-1}\left(  k\right)  \in\mathcal{C}$ for $k\in K$

(after noting that the local game induced at $k$ by $\left(  \psi
,\lambda\right)  $ is $\lambda\left(  k\right)  w_{S\left(  k\right)  }$).
This proves (i).

Next if $\mathbb{R}^{\mathcal{C}}\supsetneqq Sp\left(  \mathcal{C}\right)  $
then there exists $w^{\prime}\in Sp\left(  \mathcal{C}\right)  $ with two
different linear expansions in terms of the $MM$ games $\left\{  w_{S}%
:S\in\mathcal{C}\right\}  $ and two different representations corresponding to
them whose equitable solutions are also different (as the reader may easily
check). This proves the implication $\Longrightarrow$ of (ii). The reverse
implication follows from Theorem \ref{general TU}
\end{proof}

\end{document}